\documentclass[aps,physrev,twocolumn,groupedaddress,nofootinbib]{revtex4-2}

\usepackage{comment}
\usepackage{graphicx}
\usepackage[version=4]{mhchem}
\usepackage[colorlinks=true, linkcolor=blue, citecolor=blue, urlcolor=blue]{hyperref}
\usepackage{graphicx}

\usepackage{nasa}

\begin{document}


\title{Paleodetectors for neutrino signals from diverse Galactic stellar collapses}


\author{Mahiro Yamasaki}
\affiliation{Department of Physics, Kyushu University, Fukuoka 819-0395, Japan}

\author{Ken'ichiro Nakazato}
\affiliation{Faculty of Arts and Science, Kyushu University, Fukuoka 819-0395, Japan}



\begin{abstract}
The detectability of neutrinos from past Galactic core collapse supernovae (SNe) is investigated in terms of nuclear recoil tracks in ancient minerals, known as paleodetectors. To account for the diversity of core-collapse outcomes, variations in neutron star (NS) masses as well as failed SNe leading to black hole (BH) formation are taken into account. The role of the nuclear equation of state (EOS) is also considered, as it determines NS radii and maximum masses. The enhancement of sensitivity is quantified for models with larger neutrino emission. This emission reflects the gravitational energy released during core collapse and is larger for more compact remnants in the NS-forming case and for EOS with larger maximum masses in the BH-forming case. Furthermore, motivated by the hypothesis that nearby SN activity may have contributed to the snowball Earth events, the sensitivity to SNe occuring in a burst-like manner is also investigated. The results indicate that burst-like SN activity involving several tens of events at a distance of 10 pc would be within the reach of paleodetectors.
\end{abstract}


\maketitle

\section{Introduction}
\label{sec:intro}

Supernova (SN) explosions are among the most outstanding events in the Galaxy. Despite their low occurrence rate of only a few per century \cite{2011MNRAS.412.1473L,2013ApJ...778..164A,2021NewA...8301498R,2025MNRAS.538.1367Q}, the Galaxy has experienced numerous SN explosions over its more than 13-Gyr history and the Galactic chemical evolution is driven by these events through the ejection of heavy elements synthesized in stars. Except for type Ia SNe, SN explosions are triggered by gravitational collapse of massive stars with masses exceeding $8M_\odot$ and lifetimes shorter than 50 Myr \cite{2002RvMP...74.1015W,2009MNRAS.395.1409S,2017A&A...601A..29Z}. As this timescale is much shorter than the age of the Galaxy, core collapse (CC) SNe are associated with actively star-forming regions and young stars.

To investigate the Galactic star formation history and the CC SN rate, Baum et al. \cite{2020PhRvD.101j3017B} proposed the use of paleodetectors. This experimental concept involves searching for traces of recoiling nuclei preserved in ancient minerals with ages of order 1 Gyr \cite{2023PDU....4101245B}. From CC SNe, copious neutrinos are emitted, inducing nuclear recoils in paleodetectors. In this pioneering study \cite{2020PhRvD.101j3017B}, the minimum detectable CC SN rate averaged over the mineral age was evaluated, along with the sensitivity to time- and space-localized enhancements in the local CC SN rate motivated by starburst events.

Past SN explosions would have influenced not only Galactic evolution but also the history of the solar system and the Earth. In particular, the Earth may have been exposed to cosmic rays from a SN that occurred near the solar system. If the solar system encountered a young massive cluster, which is a dense aggregate of young stars, the Earth could have experienced substantial irradiation from sporadic bursts of CC SNe within the cluster. Indeed, young massive clusters with masses exceeding $10^4M_\odot$ and radii of order 1 pc are observed in the spiral arms of the Galaxy \cite{2010ARA&A..48..431P,2019ARA&A..57..227K}. Although no such young massive clusters have been found in the vicinity of the solar system at present, studies of Galactic chemical evolution suggest that the solar system has migrated within the Galaxy since its birth \cite{1996A&A...314..438W,2012A&A...539A.143N,2020ApJ...904..137T,2024ApJ...976L..29B}.

According to Tsujimoto \& Baba \cite{2020ApJ...904..137T}, the Sun’s birthplace was located deeper within the Galactic disk (3--5 kpc from the Galactic center) than its current position (8.2 kpc). Based on numerical simulations of the dynamical evolution of disk stars in a galaxy modeled after the Milky Way, they showed that radial migration, such as that likely experienced by the solar system, is induced by several major encounters with spiral arms. They further suggested that these spiral-arm encounters may be connected to snowball Earth events, periods during which all liquid water on the Earth’s surface was frozen \cite{2019AREPS..47....1H}. Spiral arms host many young massive stars and therefore numerous CC SNe. The enhanced influx of cosmic rays to the Earth is hypothesized to increase cloudiness, leading to strong cooling \cite{2002PhRvL..89e1102S,2007A&G....48a..18S,2017NatCo...8.2199S}.

If snowball Earth events are triggered by CC SNe in the solar neighborhood, they are potential targets for paleodetectors. For this purpose, we utilize the results of Baum et al. \cite{2020PhRvD.101j3017B}, scaled to the present situation. Accordingly, under the assumption that epsomite samples with the uranium-238 concentration of $C_{238}=0.01$~ppb are employed, a detectable signature would be left in paleodetectors if more than 50 CC SNe occurred in a burst-like manner at a distance of 10 pc from the Earth between 0.6 and 0.8 Gyr ago, encompassing the epoch of snowball Earth events \cite{2019AREPS..47....1H}. In such a scenario, young massive clusters could plausibly host more than 50 CC SNe. The expected number of CC SNe, $N_{\rm SN}$, is related to the total stellar mass of a cluster, $M_{\rm cluster}$, through $N_{\rm SN}=k_{\rm SN}M_{\rm cluster}$, where $k_{\rm SN}$ denotes the number of massive stars that explode as CC SNe per unit mass. While $k_{\rm SN}$ depends on the stellar initial mass function and the progenitor mass range of CC SNe, it is of order $0.01M_\odot^{-1}$ \cite{2014ARA&A..52..415M,2021ApJ...909..169K,2022ApJ...937...30A,2023ApJ...953..151A}. For young massive clusters with $M_{\rm cluster}>10^4M_\odot$, $N_{\rm SN} \gtrsim 100$ is therefore expected.

As described above, paleodetectors offer a promising approach for probing Galactic CC SN activity on geological timescales, which may have influenced the history of the Earth. Furthermore, Ref.~\cite{2022PhRvD.106l3008B} investigated the potential of combining paleodetector data with future observations of the diffuse SN neutrino background to infer the flavor dependence of CC SN neutrino spectra. However, the diversity of stellar core-collapse events and the uncertainties in the neutrino spectra emitted from CC SNe were not taken into account in the previous study \cite{2020PhRvD.101j3017B,2022PhRvD.106l3008B}.

Some massive stars undergo core collapse to form black holes (BHs) without producing a SN explosion; these events are referred to as failed SNe. Notably, recent observations in the nearby Andromeda galaxy (M31) have reported candidate failed SNe, where massive stars disappear without a bright SN explosion, consistent with direct BH formation \cite{2026Sci...391..689D}. Even in such cases, a large number of neutrinos are emitted, similar to CC SNe \cite{2004ApJS..150..263L,2006PhRvL..97i1101S,2020PhRvD.101l3013W,2025OJAp....8E.167S}. For ordinary CC SNe, the total energy of the emitted neutrinos corresponds to the binding energy of the neutron star (NS) left after the explosion and therefore depends on its mass and radius. Observations have revealed a population of NSs with relatively large masses \cite{2018MNRAS.478.1377A,2023Univ...10....3R,2024PhRvD.109d3052F,2025PhRvD.111b3029G,2025PhRvD.112b3045B}. The NS radius is determined by the nuclear equation of state (EOS), which remains theoretically uncertain \cite{2016ARA&A..54..401O,2021ARNPS..71..433L,2024PhRvD.110j3040L,2025PhRvX..15b1014K}. The nuclear EOS also governs neutrino emission from failed SNe \cite{2007ApJ...667..382S,2013ApJ...774...17S,2020ApJ...894....4D,2025OJAp....8E.167S}.

In the present study, we quantify how uncertainties in the nuclear EOS and the diversity of core-collapse outcomes, including both ordinary and failed SNe, affect the sensitivity of paleodetectors to the time-averaged CC SN rate in the Galaxy. We also assess their impact on the detectability of CC SNe occurring in a burst-like manner, taking a possible connection to snowball Earth events as a reference case. In both investigations, we adopt the calculation scheme developed in the preceding study \cite{2020PhRvD.101j3017B,thomas_edwards_2019_3245799}. This paper is organized as follows. In Sec.~\ref{sec:formulation}, we describe the track-length spectra in paleodetectors arising from nuclear recoils induced by neutrinos from Galactic core-collapse events, together with the assumed neutrino emission spectra from ordinary and failed SNe for various EOSs. In Sec.~\ref{sec:results}, we present the sensitivity results of paleodetectors to both the time-averaged Galactic CC SN rate and burst-like CC SN events. We devote Sec.~\ref{sec:conclusion} to the conclusions.

\section{Neutrino signals in paleodetectors} \label{sec:formulation}
In this section, we describe the evaluation method for the track-length spectra induced in paleodetectors. Except for contributions from neutrinos associated with Galactic core-collapse events, the setup follows that of the preceding study \cite{2020PhRvD.101j3017B}. The same applies to the assumptions regarding background contributions. Although paleodetectors were originally proposed as a method for the direct detection of dark matter \cite{2019PhRvD..99d3014D,2019PhRvD..99d3541E,2020PhLB..80335325B,2021PhRvD.104l3015B}, such contributions are not taken into account here.

\subsection{Neutrino spectra from stellar core-collapse events}
\label{sbsec:spectra}

As mentioned above, the neutrino spectra emitted from stellar core-collapse events depend on the remnant and the nuclear EOS. In the previous work \cite{2020PhRvD.101j3017B}, supernova neutrinos were described using a single spectral model based on a pinched Fermi–Dirac distribution \cite{2003ApJ...590..971K}. In this study, we utilize the neutrino spectra presented in Ref.~\cite{2022ApJ...937...30A} to evaluate the sensitivity of paleodetectors. These spectra are derived from numerical simulations of stellar core collapse and proto--NS cooling \cite{2021PASJ...73..639N,2022ApJ...925...98N}, employing three nuclear EOS models: Togashi \cite{2017NuPhA.961...78T}, LS220 \cite{1991NuPhA.535..331L} and Shen \cite{2011ApJS..197...20S}. Furthermore, Ref.~\cite{2022ApJ...937...30A} provides neutrino spectra for different core-collapse outcomes, including models that leave behind NSs with different masses and models in which the collapse results in BH formation without a successful SN explosion. In this study, we therefore consider three representative remnant types: a canonical-mass NS (CNS), a high-mass NS (HNS), and a failed SN forming a BH. The gravitational masses of the CNS and HNS are assumed to be approximately $1.34M_\odot$ and $1.65M_\odot$, respectively.

The properties of the neutrino spectra for each model are summarized in Table~1 of Ref.~\cite{2022ApJ...937...30A}. For CC SNe, the total energy of the emitted neutrinos is determined by the gravitational binding energy of the remnant NS, leading to larger emission energies in the HNS models than in the CNS models. Since the binding energy scales approximately with the inverse of the NS radius, models predicting smaller radii result in larger total neutrino energies. Accordingly, among the three EOS models adopted in this study, the Togashi EOS produces the most compact NSs and hence the largest total neutrino energy. For failed SNe, mass accretion onto the proto--NS and associated heating continue until just before BH formation. As a result, the average energies of the emitted neutrinos are higher than those from ordinary CC SNe. The duration and total output of neutrino emission are governed by the maximum mass of the proto--NSs, which is determined by the nuclear EOS. Since the Shen EOS predicts the highest maximum mass, it results in the longest neutrino emission timescale and hence the largest total neutrino energy among the three EOS models.

In Ref.~\cite{2022ApJ...937...30A}, the spectra of $\nu_e$, $\bar\nu_e$, and $\nu_x$ are given separately, where $\nu_x$ is defined as the average over $\nu_\mu$, $\bar\nu_\mu$, $\nu_\tau$, and $\bar\nu_\tau$. In the present study, however, we focus on neutral-current interactions between neutrinos and nuclei, which are flavor independent, and on the damage tracks in paleodetectors produced by recoiling nuclei. Thus, hereafter we consider only the neutrino spectra summed over all flavors,
\begin{equation}
\left(\frac{\mathrm{d}N}{\mathrm{d}E_{\nu}}\right)
=
\left(\frac{\mathrm{d}N}{\mathrm{d}E_{\nu}}\right)_{\nu_e}
+
\left(\frac{\mathrm{d}N}{\mathrm{d}E_{\nu}}\right)_{\bar{\nu}_e}
+4
\left(\frac{\mathrm{d}N}{\mathrm{d}E_{\nu}}\right)_{\nu_x},
\label{eq:dnde_sn}
\end{equation}
for each model characterized by a given remnant and EOS. Here, $E_{\nu}$ denotes the neutrino energy.

\subsection{Neutrino flux from Galactic core collapses at the Earth}
\label{sbsec:flux}
To evaluate the spectrum of time-averaged neutrino flux from Galactic core-collapse events at Earth, the spatial distribution of such events is required. In this study, we adopt the Galactic CC SN distribution model of Ref.~\cite{2013ApJ...778..164A}, following the approach of Ref.~\cite{2020PhRvD.101j3017B}. The distribution is expressed in a double-exponential form as a function of the galactocentric radius $R$ and the height $z$ above the Galactic midplane,
\begin{equation}
\rho(R,z) \propto \exp\left(-\frac{R}{R_d}\right)\exp\left(-\frac{|z|}{H}\right)
\label{eq:dist_ccsn}
\end{equation}
with $R_d=2.9$~kpc and $H=95$~pc. Then, we adopt a galactocentric radius of $R_0 = 8.7$~kpc and a height of $z_0 = 24$~pc for the present Solar position, noting that these parameters are subject to uncertainties in the literature. In particular, since the adopted value $R_0 = 8.7$~kpc \cite{2009A&A...498...95V} is relatively large compared to other estimates \cite{2009MNRAS.398..263M,2014MNRAS.441.1105F,2019A&A...625L..10G,2020PASJ...72...50V}, it is expected that the resulting time-averaged neutrino flux is conservative. Furthermore, the Solar coordinates are taken to be constant throughout the exposure period relevant for paleodetectors. Although numerical simulations indicate possible Solar radial migration of $\lesssim 1$~kpc over the past $\sim 1$~Gyr~\cite{2020ApJ...904..137T}, such effects on the time-averaged neutrino flux are neglected for simplicity and consistency with Ref.~\cite{2020PhRvD.101j3017B}. The influence of radial migration, as well as more complex Galactic structures such as spiral arms, is deferred to future studies.

Based on the adopted spatial distribution of Galactic CC SNe $\rho(R,z)$ and the Solar position $(R_0,z_0)$, we calculate the probability density $f(R_{\mathrm{E}})$ for a CC SN to occur at a distance $R_{\mathrm{E}}$ from Earth. By integrating over $f(R_{\mathrm{E}})$, we obtain the spectrum of time-averaged neutrino flux from Galactic CC SNe at Earth as
\begin{equation}
\left(\frac{\mathrm{d}\Phi}{\mathrm{d}E_{\nu}}\right)_{\rm SN}
=
R_{\rm SN}
\left(\frac{\mathrm{d}N}{\mathrm{d}E_{\nu}}\right)_{\rm SN}
\int \frac{f(R_{\rm E})}{4\pi R_{\rm E}^{2}} \,\mathrm{d}R_{\rm E},
\label{eq:flux_sn}
\end{equation}
where $R_{\mathrm{SN}}$ is the Galactic CC SN rate, including contributions from both CNSs and HNSs. The neutrino spectrum per SN is given by
\begin{align}
\left(\frac{\mathrm{d}N}{\mathrm{d}E_{\nu}}\right)_{\rm SN}
&= (1-f_{\rm HNS})
   \left(\frac{\mathrm{d}N}{\mathrm{d}E_{\nu}}\right)_{\rm CNS} \notag\\
&\quad + f_{\rm HNS}
   \left(\frac{\mathrm{d}N}{\mathrm{d}E_{\nu}}\right)_{\rm HNS},
\label{eq:dnde_sn_total}
\end{align}
with $(\mathrm{d}N/\mathrm{d}E_{\nu})_{\rm CNS}$ and $(\mathrm{d}N/\mathrm{d}E_{\nu})_{\rm HNS}$ representing the neutrino spectra for SNe forming CNSs and HNSs, respectively. While the mass distribution of NSs at birth remains uncertain, we parametrize this uncertainty solely by the fraction of SNe forming HNSs, $f_{\rm HNS}$, as adopted in Ref.~\cite{2022ApJ...937...30A}.

Recent Bayesian analyses of NS observations, which simultaneously infer the nuclear EOS and the NS mass distribution, suggest a bimodal structure in the NS mass population \cite{2018MNRAS.478.1377A,2023Univ...10....3R,2024PhRvD.109d3052F,2025PhRvD.111b3029G,2025PhRvD.112b3045B}. The low-mass component is characterized by a narrow peak at $1.33$--$1.34M_\odot$, whereas the high-mass component exhibits a broad distribution spanning $1.5$--$1.9M_\odot$. These Bayesian mixture models carry a weight of approximately $0.32$--$0.44$ for the high-mass component. This fraction should be considered an upper limit on our $f_{\rm HNS}$, corresponding to the intrinsic birth fraction, because subsequent mass accretion from binary companions can increase NS masses and thereby enhance the apparent high-mass population in present-day observations. Indeed, a much smaller birth fraction of high-mass NSs is inferred from stellar evolution models \cite{2020ApJ...896...56W}. 

One of the new elements of this work is the inclusion of contributions from failed SNe leading to BH formation in the analysis of neutrino signals in paleodetectors. In analogy with Eq.~(\ref{eq:flux_sn}) for ordinary SNe, the spectrum of the time-averaged neutrino flux from Galactic failed SNe at Earth is expressed as
\begin{equation}
\left(\frac{\mathrm{d}\Phi}{\mathrm{d}E_{\nu}}\right)_{\rm BH}
=
R_{\rm BH}
\left(\frac{\mathrm{d}N}{\mathrm{d}E_{\nu}}\right)_{\rm BH}
\int \frac{f(R_{\rm E})}{4\pi R_{\rm E}^{2}} \,\mathrm{d}R_{\rm E},
\label{eq:flux_bh}
\end{equation}
where $(\mathrm{d}N/\mathrm{d}E_{\nu})_{\rm BH}$ represents the neutrino spectrum for failed SNe. In this treatment, we assume that the spatial probability density $f(R_{\mathrm{E}})$ for a failed SN to occur at a distance $R_{\mathrm{E}}$ from Earth is identical to that adopted for a CC SN. Since the Galactic failed SN rate $R_{\mathrm{BH}}$ is intrinsically difficult to constrain observationally, we instead focus on the fraction of failed SNe among all core-collapse events, defined as
\begin{equation}
f_{\rm{BH}} = \frac{R_{\rm{BH}}}{R_{\rm{SN}} + R_{\rm{BH}}}.
\label{eq:f_bh}
\end{equation}
Failed SNe associated with BH formation are expected to manifest observationally as the disappearance of luminous stars without accompanying SN explosions at optical wavelengths \cite{2008ApJ...684.1336K}. An observational constraint on such events was obtained in Ref.~\cite{2021MNRAS.508..516N}, which compiled 11 years of monitoring data for luminous stars in nearby galaxies using the Large Binocular Telescope. Provided that one failed SN and eight successful CC SNe were detected, the fraction of failed SNe among all core-collapse events was estimated to be $f_{\mathrm{BH}} = 0.04$--$0.39$ at the 90\% confidence level. This range is broadly consistent with predictions from stellar evolution models, which suggest $f_{\mathrm{BH}} \simeq 0.09$--$0.32$ \cite{2020ApJ...896...56W}.

In this study, we treat $f_{\rm HNS}$ and $f_{\rm BH}$ as free parameters and calculate the spectrum of the total neutrino flux from Galactic core-collapse events as
\begin{equation}
\left(\frac{\mathrm{d}\Phi}{\mathrm{d}E_{\nu}}\right)_{\rm{total}} = \left(\frac{\mathrm{d}\Phi}{\mathrm{d}E_{\nu}}\right)_{\rm{SN}}+\left(\frac{\mathrm{d}\Phi}{\mathrm{d}E_{\nu}}\right)_{\rm{BH}},
\label{eq:flux_total}
\end{equation}
to investigate their impact on the sensitivity of paleodetectors.

\subsection{Neutrino-induced damage track spectra in paleodetectors}
\label{sbsec:trackspectra}

Neutrinos emitted from a Galactic stellar core-collapse event can reach the Earth and produce damage tracks in minerals. Neutrinos with energies $E_{\nu} \lesssim 100\,\mathrm{MeV}$ scatter off nuclei via coherent neutral-current interactions, which are independent of the neutrino flavor. The neutrino--nucleus scattering cross section for a target nuclei is given by
\begin{equation}
\begin{aligned}
\left(\frac{\mathrm{d}\sigma}{\mathrm{d}E_{\rm{R}}}\right)_{\mathrm{T}}
(E_{\rm{R}},E_{\nu})
&= \frac{G_{\rm{F}}^2}{4\pi} Q_{\rm{W}}^2 m_{\rm{T}} \\
&\quad \times \left(1-\frac{m_{\rm{T}}E_{\rm{R}}}{2E_{\nu}^2}\right)
F^2(E_{\rm{R}}),
\end{aligned}
\label{eq:dsigmade}
\end{equation}
where the index $\mathrm{T}$ labels each target nucleus contained in minerals used as paleodetectors. In this expression, $m_{\mathrm{T}}$ is the mass of $\mathrm{T}$, $E_{\rm{R}}$ is the nuclear recoil energy, and $G_{\rm{F}}$ is the Fermi coupling constant. The weak charge $Q_{\mathrm{W}}$ depends on the number of nucleons $A_{\mathrm{T}}$ and protons $Z_{\mathrm{T}}$ in the target nucleus as
\begin{equation}
Q_{\rm{W}}=(A_{\rm{T}}-Z_{\rm{T}})-(1-4\sin^2{\theta_{\rm{W}}})Z_{\rm{T}},
\label{eq:Q_W}
\end{equation}
where $\theta_{\mathrm{W}}$ is the weak mixing angle. The nuclear form factor $F(E_{\mathrm{R}})$ encodes the finite-size structure of the nucleus. In the present analysis, we adopt the Helm form factor, given by
\begin{equation}
F(E_{\rm{R}})=3\frac{\sin{(qr_n)}-qr_n\cos{(qr_n)}}{{(qr_n)}^3}e^{{-(qs)}^2/2}
\label{eq:form_factor}
\end{equation}
where $q=\sqrt{2m_{\mathrm{T}}E_{\mathrm{R}}}$ is the momentum transfer. The effective nuclear radius $r_n$ is taken to be $r_n^2={c}^2+\frac{7}{3}{\pi}^2{a}^2-5s^2$ with $a=0.52 \, \mathrm{fm}$, $c=(1.23{A_{\rm{T}}^{1/3}}-0.6) \, \mathrm{fm}$, and $s=0.9 \,\mathrm{fm}$.

For a target nucleus $\mathrm{T}$, the differential recoil spectrum per unit target mass is derived by integrating over the neutrino energy as
\begin{equation}
\left(\frac{\mathrm{d}R}{\mathrm{d}E_{\rm{R}}}\right)_{\rm{T}}=\frac{1}{m_{\rm{T}}}\int\left(\frac{\mathrm{d}\sigma}{\mathrm{d}E_{\rm{R}}}\right)_{\rm{T}}\left(\frac{\mathrm{d}\Phi}{\mathrm{d}E_{\nu}}\right)_{\rm{total}}\mathrm{d}E_{\nu}.
\label{eq:dedr}
\end{equation}
The recoil energy $E_{\mathrm{R}}$ is mapped onto the damage track length $x_{\mathrm{T}}$ via
\begin{equation}
x_{\mathrm{T}}(E_{\rm{R}}) = \int^{E_{\rm{R}}}_{0} \mathrm{d}E \, \left | \frac{\mathrm{d}E}{\mathrm{d}x_{\mathrm{T}}} \right|^{-1},
\label{eq:x}
\end{equation}
where the stopping power $\left|\mathrm{d}E/\mathrm{d}x_{\mathrm{T}}\right|$ represents the energy loss per unit path length. The total damage track-length spectrum is calculated by converting the recoil energy to track length and summing over all nuclear species,
\begin{equation}
\frac{\mathrm{d}R}{\mathrm{d}x}=\sum_{\rm{T}} \, \xi_{\rm{T}} \, \frac{\mathrm{d}E_{\rm{R}}}{\mathrm{d}x_{\rm{T}}} \, \left(\frac{\mathrm{d}R}{\mathrm{d}E_{\rm{R}}}\right)_{\rm{T}},
\label{eq:drdx}
\end{equation}
where $\xi_{\mathrm{T}}$ denotes the mass fraction of nucleus $\mathrm{T}$ in the target mineral. In this study, the evaluation of the stopping power is obtained from \texttt{SRIM} code \cite{2010NIMPB.268.1818Z}, as adopted in Ref.~\cite{2020PhRvD.101j3017B}. Recent Monte Carlo studies have reported a modification of the track-length spectrum due to electronic stopping power, characterized by a suppression of tracks with lengths exceeding $x \sim 200\,\mathrm{nm}$ and a hump-like feature around this scale~\cite{2025PhRvD.112d3040F}. For simplicity, this effect is not taken into account in the present analysis.

\subsection{Background assumptions}
\label{sbsec:background}

Since the observables of paleodetectors are the damage tracks, any particles that produce damage tracks similar to those induced by neutrinos are regarded as background. We briefly summarize the background components following Ref.~\cite{2020PhRvD.101j3017B}.

Radioactive elements in paleodetectors constitute an important source of background, with uranium-238 being the dominant contributor. Fast neutrons are generated both by spontaneous fission of uranium-238 and by $(\alpha,n)$-reactions induced by $\alpha$ particles from its decay chain. These neutrons elastically scatter off nuclei in the mineral and produce damage tracks with lengths comparable to those induced by neutrinos. The neutron spectra of spontaneous fission and $(\alpha,n)$ reactions are taken from calculations based on the \texttt{SOURCES-4A} code \cite{Madland1999-ee}, and the corresponding nuclear recoil spectra are evaluated based on the \texttt{JANIS4.0} database \cite{2014NDS...120..294S}. The decay of uranium-238 proceeds through a full decay chain involving eight $\alpha$ decays, producing a set of eight spatially connected tracks that are distinguishable from isolated neutrino-induced tracks. Thus, we assume that such events can be rejected based on their characteristic track topology, except for cases in which only a single $\alpha$ decay occurs. These single-$\alpha$ events originate from ${}^{238}{\rm U} \to {}^{234}{\rm Th} + \alpha$, yielding a ${}^{234}{\rm Th}$ with a recoil energy of $72\,\mathrm{keV}$. The corresponding damage tracks have a characteristic length of $\sim 50\,\mathrm{nm}$, which are expected to have a negligible impact on the sensitivity (but see also Ref.~\cite{2025PhRvD.112d3040F}).

In addition to fast neutrons from spontaneous fission and $(\alpha,n)$ reactions, neutrons can also be produced by spallation reactions induced by cosmic-ray muons. Since such contributions are efficiently suppressed at large overburden, we assume target samples for paleodetectors to be located deep underground and neglect cosmic ray induced neutrons.

Neutrinos originating from sources other than Galactic core-collapse events also contribute to the background. In this study, we consider solar neutrinos, atmospheric neutrinos, and the diffuse SN neutrino background, which represents the cumulative flux from distant supernovae throughout cosmic history. These neutrino sources dominate different energy ranges and therefore contribute to different regions of the damage track-length spectrum. Solar neutrinos dominate short tracks ($x \lesssim 100\,\mathrm{nm}$), atmospheric neutrinos dominate long tracks ($x \gtrsim 200\,\mathrm{nm}$), and the diffuse SN neutrino background dominates the intermediate region. The fluxes of solar and atmospheric neutrinos are taken from Ref.~\cite{2016PhRvD..94f3527O}. Temporal variations in the solar and atmospheric neutrino fluxes have been examined in Refs. \cite{2021PhRvD.103l3016T} and \cite{2020PhRvL.125w1802J}, respectively, but are not included in the present analysis. Although a flux of the diffuse SN neutrino background based on the same supernova neutrino spectra has been provided in Ref.~\cite{2022ApJ...937...30A}, its flux is generally approximately two orders of magnitude smaller than the time-averaged neutrino flux from Galactic core-collapse events. Therefore, for simplicity, we adopt a single model presented in Ref.~\cite{2010ARNPS..60..439B}, as in Ref.~\cite{2020PhRvD.101j3017B}.

\begin{figure}
 \includegraphics[width=85mm]{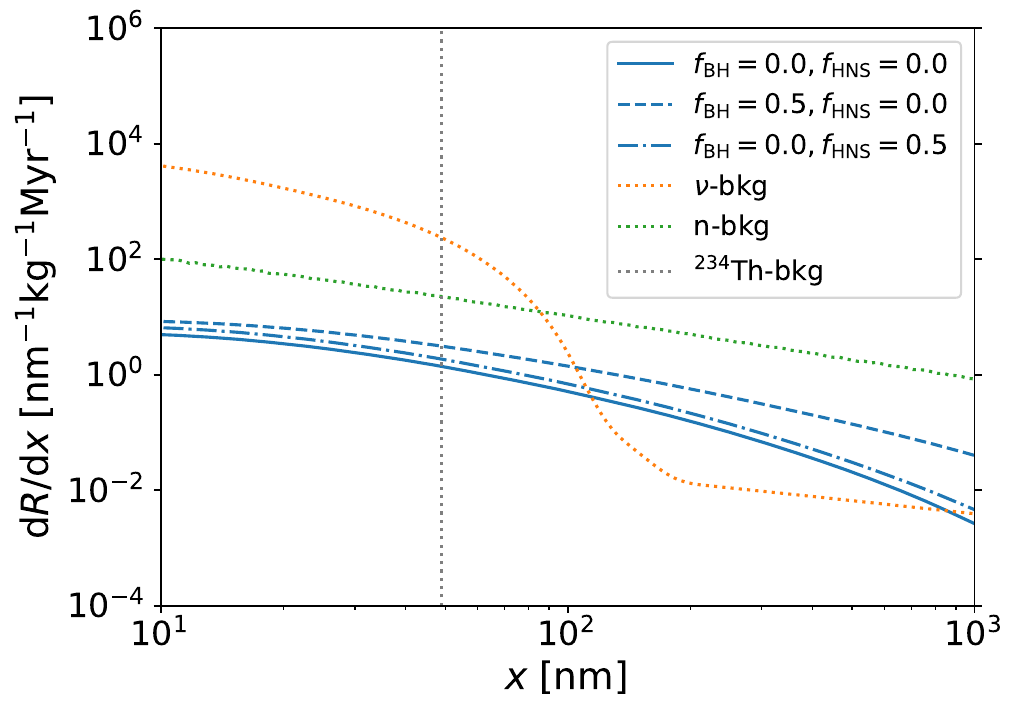}
 \caption{Damage track-length spectrum for an epsomite [$\ce{Mg(SO_4)\cdot {7}(H_2O)}$] sample with mass $M=1\,\mathrm{kg}$, age $t_{\mathrm{age}}=1\,\mathrm{Myr}$, and uranium-238 concentration $C_{238}=0.01\,\mathrm{ppb}$, assuming the Togashi EOS and a time-averaged Galactic SN rate of $R_{\mathrm{SN}}=3.0\times10^{-2}\,\mathrm{yr^{-1}}$. Note that damage track-length spectra can be estimated for arbitrary sample masses and ages by scaling linearly with these quantities. The solid, dashed, and dash-dotted lines correspond to $(f_{\mathrm{BH}}, f_{\mathrm{HNS}}) = (0.0, 0.0)$, $(0.5, 0.0)$, and $(0.0, 0.5)$, respectively. Background components are shown as dotted lines: orange, green, and gray correspond to neutrino-, neutron-, and $^{234}\mathrm{Th}$-induced background, respectively. Each background component is calculated following the settings of Ref.~\cite{2020PhRvD.101j3017B}.}
 \label{fig:drdx}
\end{figure}

In Fig.~\ref{fig:drdx}, we show the estimated damage track-length spectra for an epsomite [$\ce{Mg(SO_4)\cdot {7}(H_2O)}$] sample with a mass of $M=1\,\mathrm{kg}$, an age of $t_{\mathrm{age}}=1\,\mathrm{Myr}$ and a uranium-238 concentration of $C_{238}=0.01\,\mathrm{ppb}$, together with the background components. These parameter choices are merely illustrative, since damage track-length spectra can be obtained for arbitrary sample mass and age by scaling linearly with these quantities. Here, we adopt epsomite for the investigation, as it is the most promising among the four minerals considered in Ref.~\cite{2020PhRvD.101j3017B} due to its chemical composition. In particular, epsomite contains hydrogen, which acts as an efficient moderator of fast neutrons: since hydrogen nuclei have a mass comparable to that of neutrons, a neutron loses a large fraction of its energy in a single scattering. In addition, most nuclei in epsomite have small cross sections for $(\alpha,n)$ reactions, leading to suppression of neutron-induced backgrounds. Furthermore, epsomite is a type of marine evaporite that forms at the bottom of evaporating bodies of water. In such environments, the concentration of uranium-238 is lower by a few orders of magnitude than that in typical crustal materials. The chemical composition of epsomite also enhances the sensitivity to the signal over background in the damage track-length distribution. For nuclei lighter than $\sim 10\,\mathrm{GeV}$ (corresponding to $\ce{C}$), neutrinos from CC SNe induce damage tracks with lengths similar to those from the solar neutrino background. For heavy nuclei with the mass of $\sim 30\,\mathrm{GeV}$ (corresponding to $\ce{Si}$), on the other hand, most of the resulting tracks are too short to be resolved due to the finite readout resolution. Since most constituent nuclei in epsomite have masses between those of $\ce{C}$ and $\ce{Si}$, the detectability of the signal over background in the track-length distribution is improved. In the following, we focus on epsomite with a uranium-238 concentration of $C_{238} = 0.01\,\mathrm{ppb}$, representative of typical contamination levels in marine evaporites.

\section{Sensitivity of paleodetectors}
\label{sec:results}

In this section, we present the sensitivity of paleodetectors to neutrinos from stellar core collapses, taking into account the diversity of compact remnants and dependence on the nuclear EOS. In Sec.~\ref{sbsec:snstv_neutrino}, we first investigate the minimum Galactic CC SN rate required for detection at the $3\sigma$ level. We then evaluate the mineral age necessary to achieve detection under an assumed constant Galactic CC SN rate. In Sec.~\ref{sbsec:snstv_burst}, we explore the sensitivity to burst-like core-collapse events associated with the snowball Earth epoch.

To evaluate the sensitivity of paleodetectors, we consider an epsomite sample with a mass of $M=100\,\mathrm{g}$ and a uranium-238 concentration of $C_{238}=0.01\,\mathrm{ppb}$ as a benchmark. The damage tracks are assumed to be read out with a length resolution of $\sigma_{x}=15\,\mathrm{nm}$, motivated by the capabilities of small angle x-ray scattering tomography at synchrotron facilities \cite{2014NatSR...4.3857H}.

In this study, we employ the code developed in Ref.~\cite{2020PhRvD.101j3017B} for the sensitivity estimates, which is publicly available in Ref.~\cite{thomas_edwards_2019_3245799}. The statistical analysis is carried out using the \texttt{SWORDFISH} \texttt{PYTHON} package, which enables fast calculations of projected sensitivities \cite{2017arXiv171205401E,2018JCAP...02..021E}. The corresponding code is available online\footnote{\url{https://github.com/cweniger/swordfish}}.

\subsection{Time-averaged neutrino signal from Galactic stellar collapses}
\label{sbsec:snstv_neutrino}

\begin{figure*}
   \includegraphics[width=85mm]{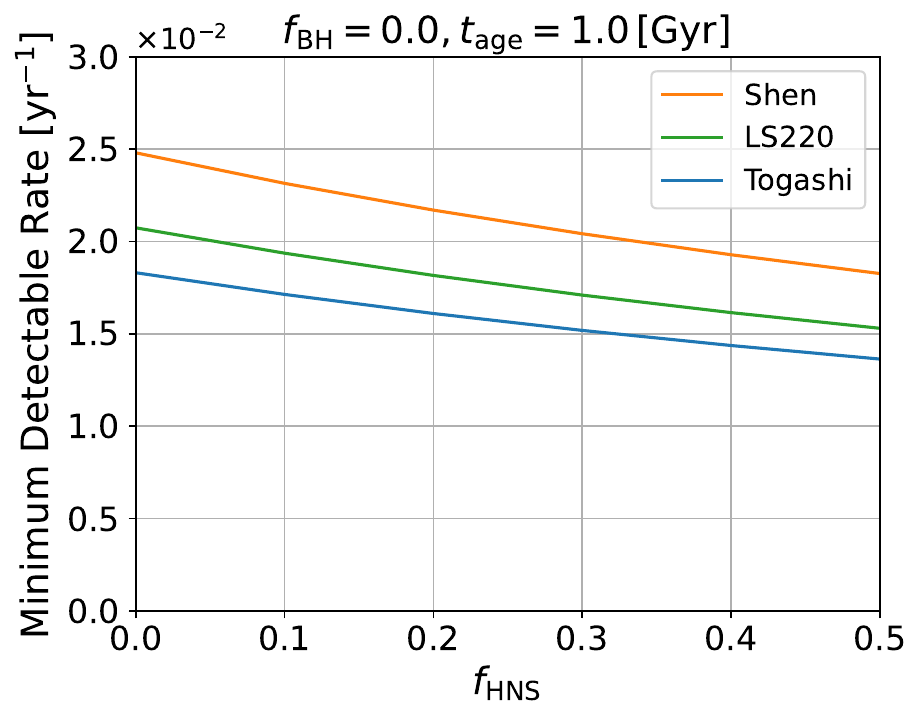}
   \includegraphics[width=85mm]{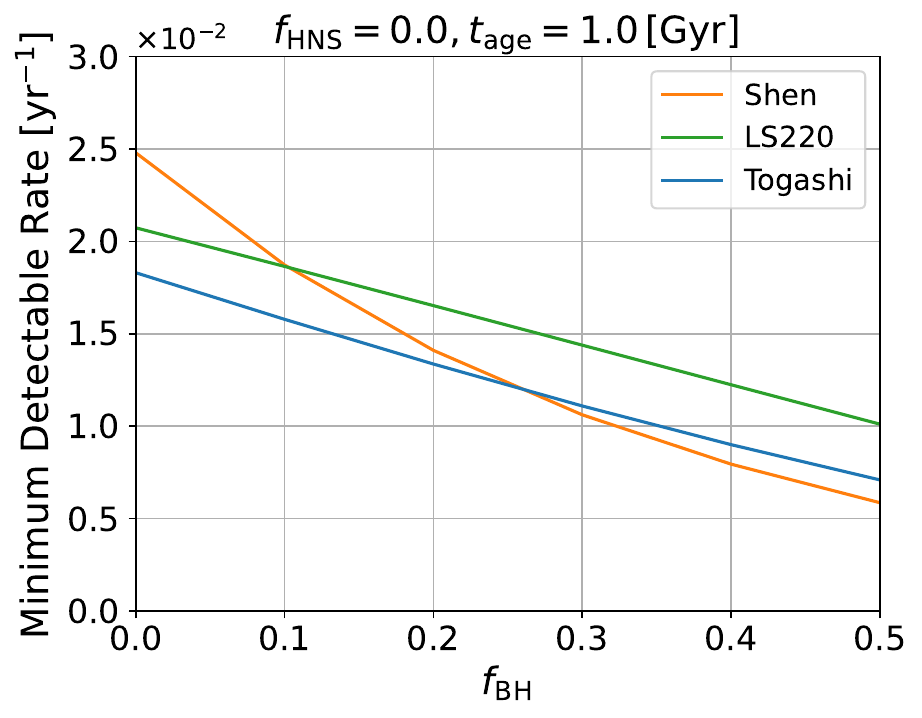}
   \caption{Left: Minimum Galactic SN rate required for neutrino detection with epsomite samples at the $3\sigma$ level as a function of $f_{\mathrm{HNS}}$ with $f_{\mathrm{BH}}=0.0$. Right: Same as the left panel, but shown as a function of $f_{\mathrm{BH}}$ with $f_{\mathrm{HNS}}=0.0$. In both panels, an epsomite sample with an age of $t_{\mathrm{age}}=1.0\,\mathrm{Gyr}$, a mass of $M=100\,\mathrm{g}$, and a uranium-238 concentration of $C_{238}=0.01\,\mathrm{ppb}$ is assumed. Blue, green, and orange lines correspond to the Togashi, LS220, and Shen EOS models, respectively.}
 \label{fig:Rmin}
\end{figure*}

\begin{figure*}[htbp]
\centering

\begin{minipage}[t]{0.32\textwidth}
    \centering
    \includegraphics[width=\textwidth]{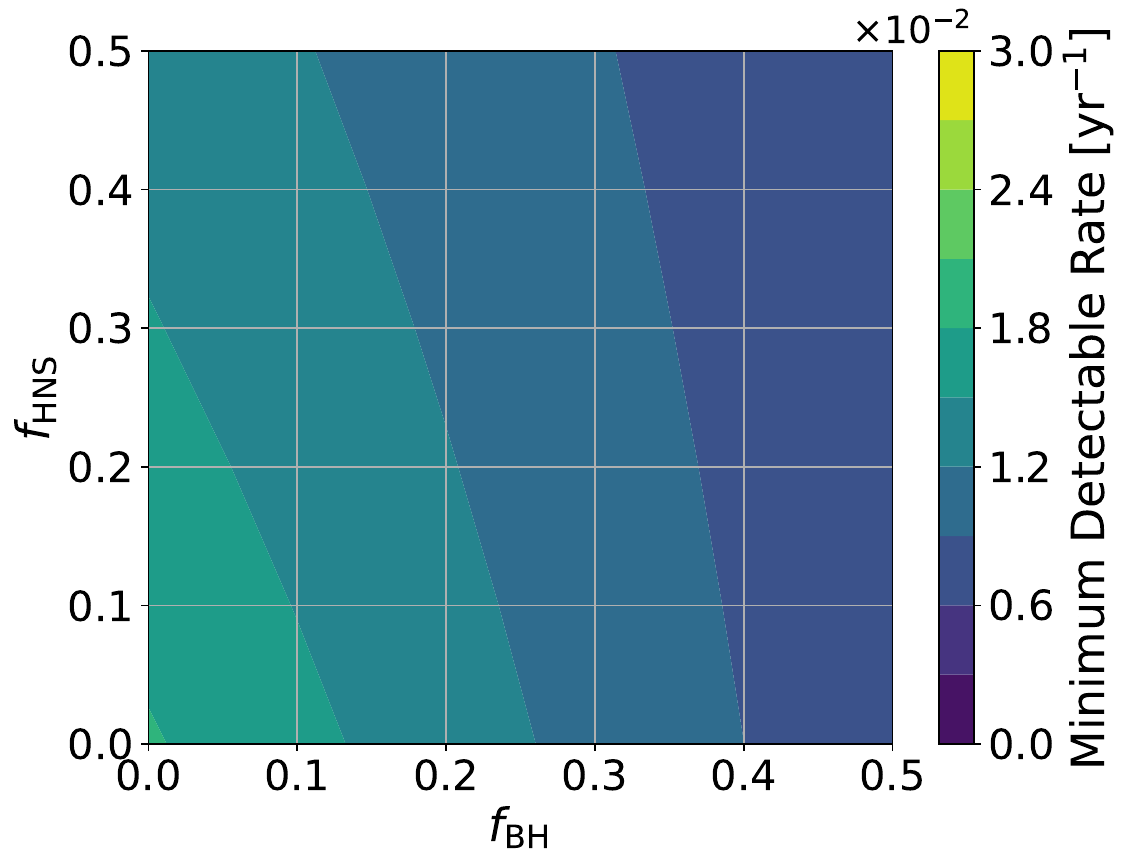} 
\end{minipage}
\hfill
\begin{minipage}[t]{0.32\textwidth}
    \centering
    \includegraphics[width=\textwidth]{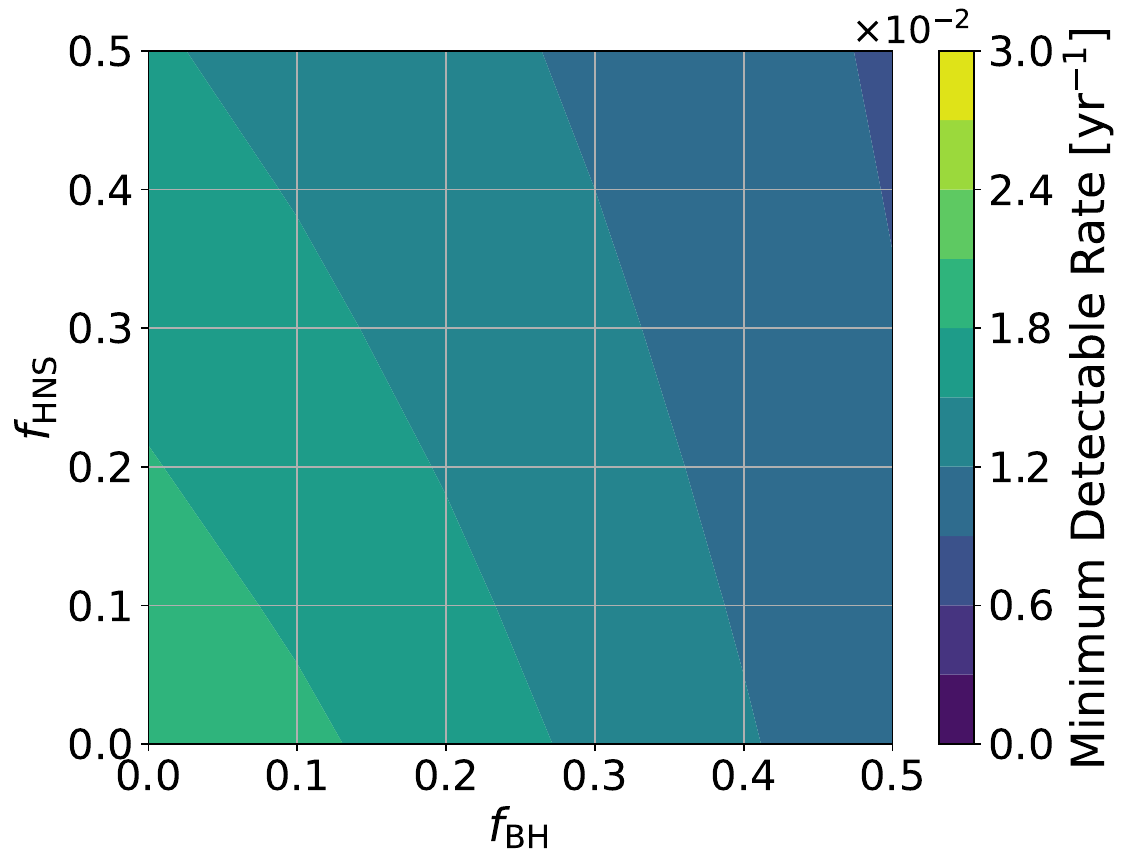}
\end{minipage}
\hfill
\begin{minipage}[t]{0.32\textwidth}
    \centering
    \includegraphics[width=\textwidth]{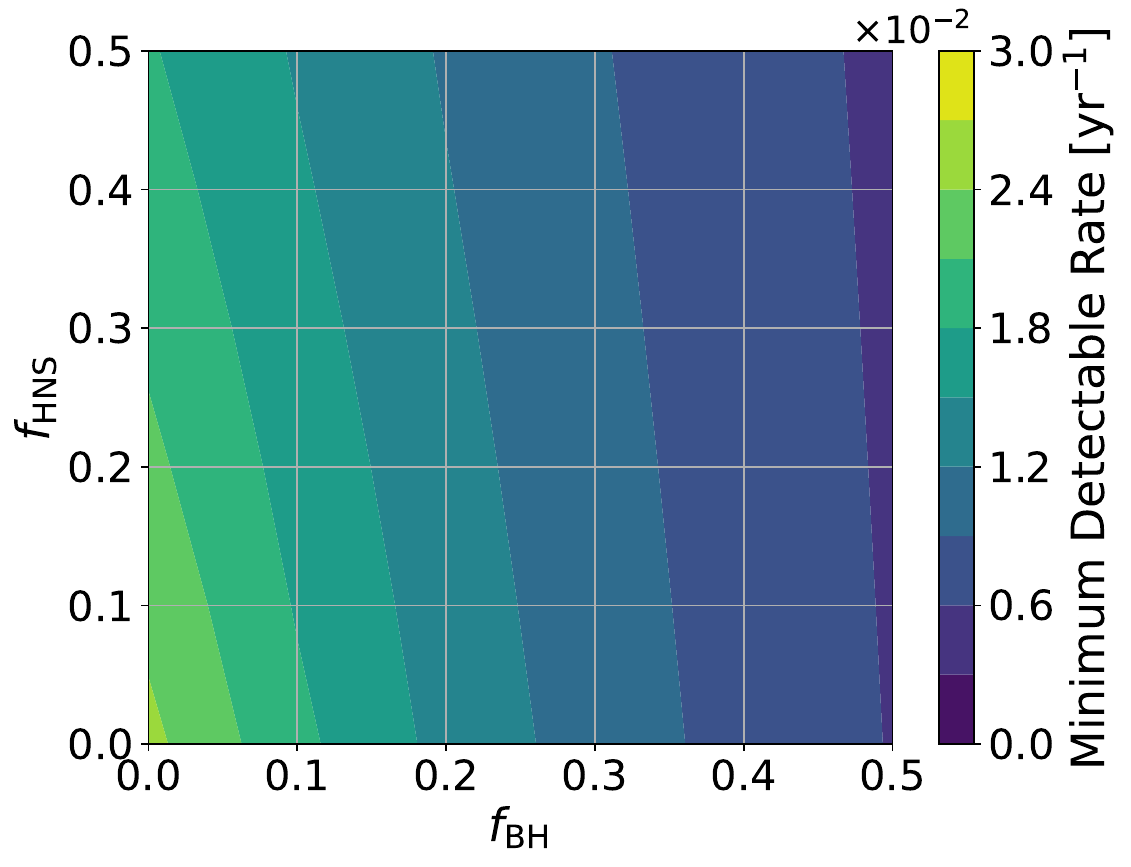}
\end{minipage}
    \caption{Minimum Galactic SN rate required for neutrino detection with epsomite samples at the $3\sigma$ level as a function of $f_{\mathrm{BH}}$ and $f_{\mathrm{HNS}}$. The assumptions for the epsomite sample are the same as in Fig.~\ref{fig:Rmin}. The left, center, and right panels show the results for the Togashi, LS220, and Shen EOS models, respectively.}
    \label{fig:R_min_contour}
\end{figure*}

\begin{figure*}
   \includegraphics[width=85mm]{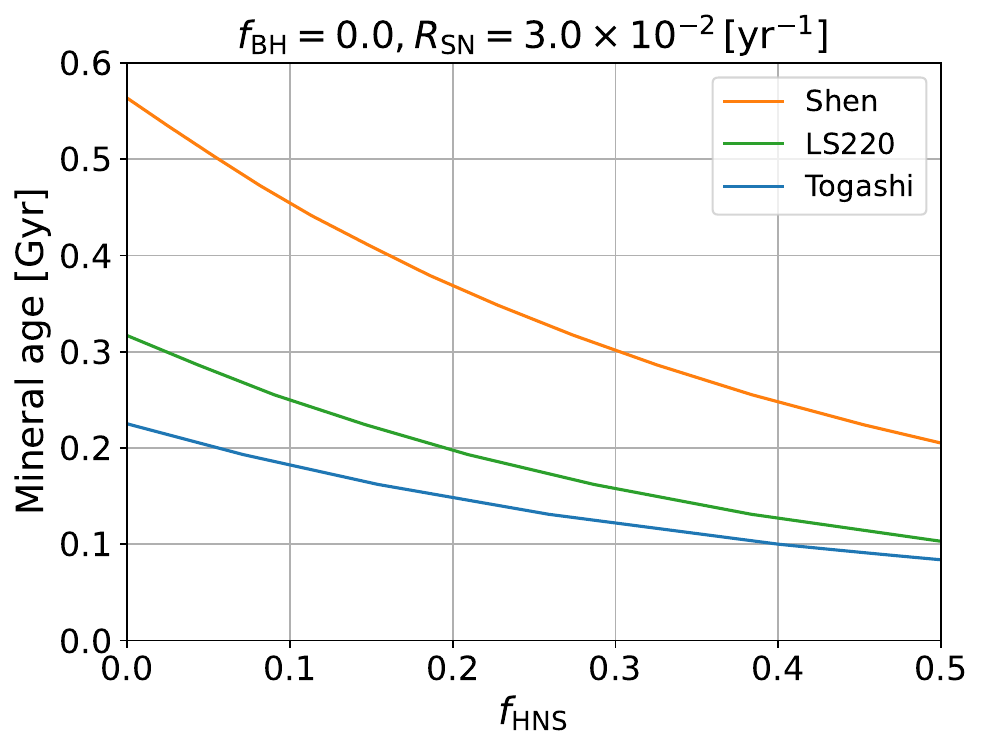}
   \includegraphics[width=85mm]{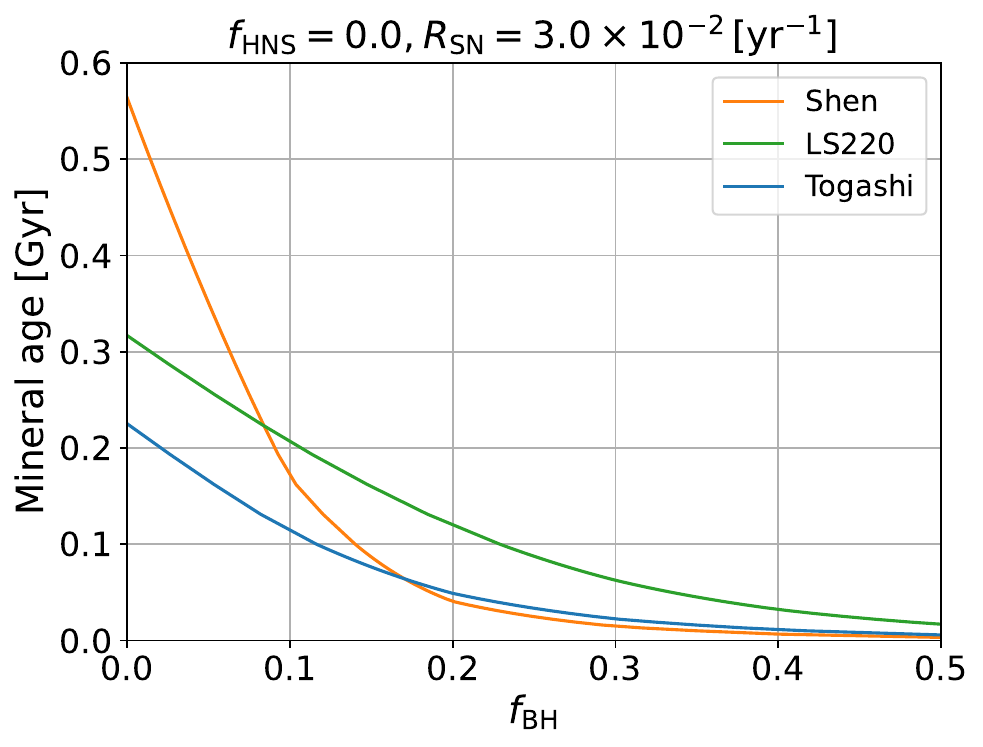}
   \caption{Left: Mineral age of an epsomite sample required for neutrino detection at the $3\sigma$ level as a function of $f_{\mathrm{HNS}}$ with $f_{\mathrm{BH}}=0.0$. Right: Same as the left panel, but shown as a function of $f_{\mathrm{BH}}$ with $f_{\mathrm{HNS}}=0.0$. In both panels, we assume a time-constant Galactic SN rate of $R_{\mathrm{SN}}=3.0\times10^{-2}\,\mathrm{yr^{-1}}$ and a sample with a mass of $M=100\,\mathrm{g}$ and a uranium-238 concentration of $C_{238}=0.01\,\mathrm{ppb}$. Line colors follow Fig.~\ref{fig:Rmin}.}
 \label{fig:agemin}
\end{figure*}

\begin{figure*}[htbp]
\centering

\begin{minipage}[t]{0.32\textwidth}
    \centering
    \includegraphics[width=\textwidth]{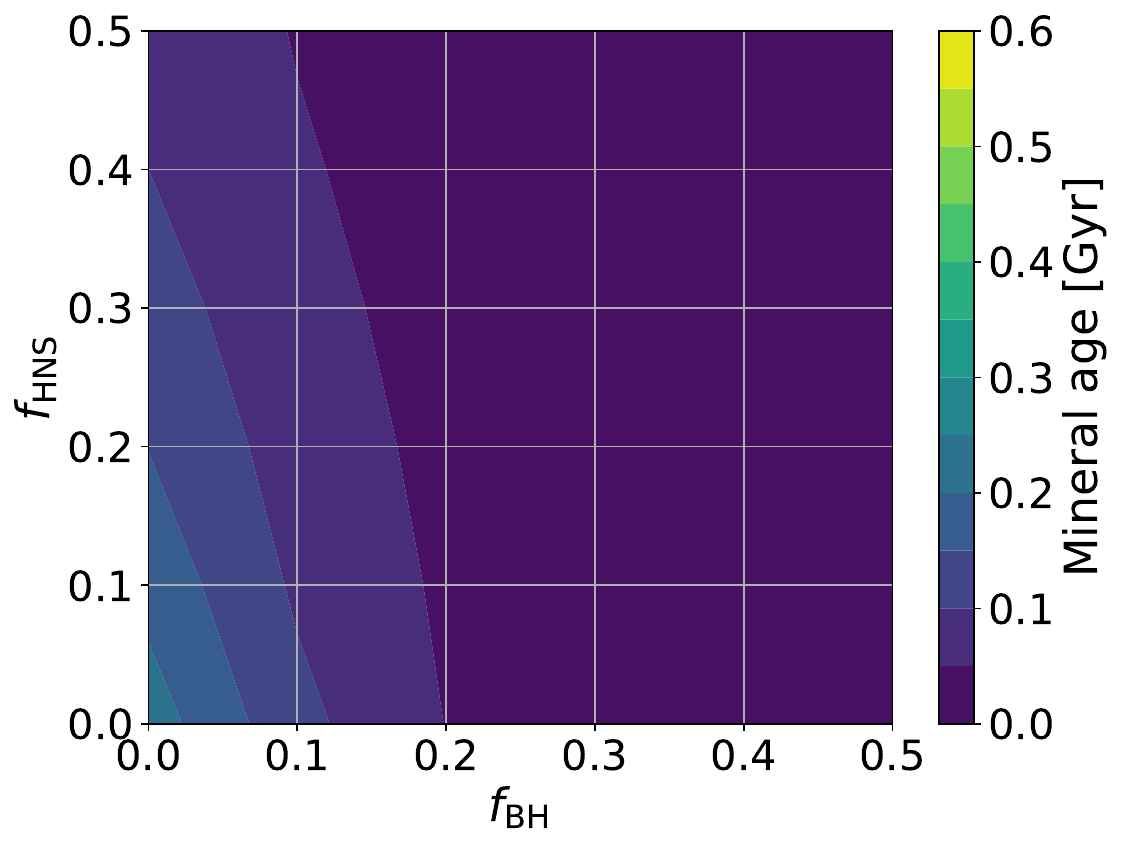}
\end{minipage}
\hfill
\begin{minipage}[t]{0.32\textwidth}
    \centering
    \includegraphics[width=\textwidth]{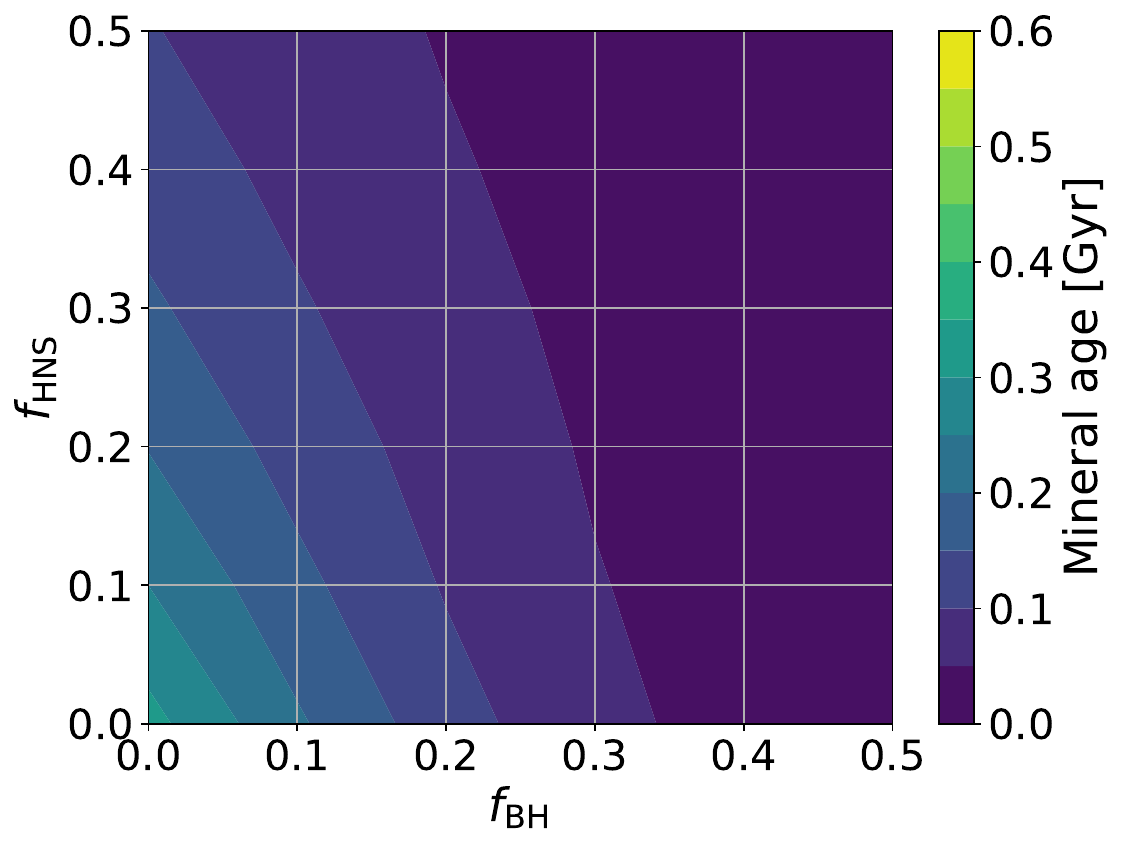}
\end{minipage}
\hfill
\begin{minipage}[t]{0.32\textwidth}
    \centering
    \includegraphics[width=\textwidth]{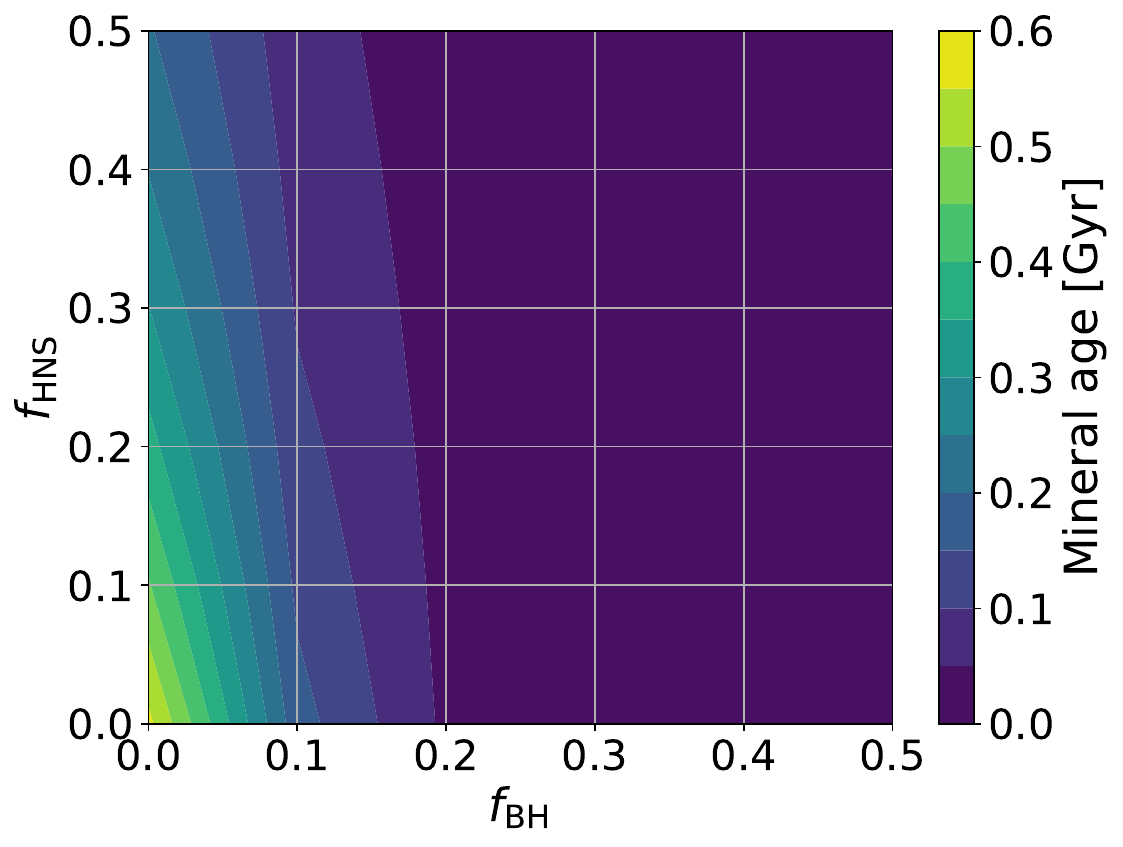}
\end{minipage}
    \caption{Mineral age of an epsomite sample required for neutrino detection at the $3\sigma$ level as a function of $f_{\mathrm{BH}}$ and $f_{\mathrm{HNS}}$. The assumptions for the Galactic SN rate and the sample are the same as in Fig.~\ref{fig:agemin}. The left, center, and right panels show the results for the Togashi, LS220, and Shen EOS models, respectively.}
\label{fig:agemin_contour}
\end{figure*}

In this subsection, we estimate the minimum Galactic CC SN rate required for a $3\sigma$ neutrino detection to quantify the sensitivity of paleodetectors. We define this rate as the time-averaged value for which the no-signal hypothesis can be rejected at the $3\sigma$ level \cite{2020PhRvD.101j3017B}. The statistical analysis follows the method used to derive the discovery reach in Ref.~\cite{2017arXiv171205401E}.

In Fig.~\ref{fig:Rmin}, we show the minimum Galactic SN rate as a function of $f_{\mathrm{HNS}}$ and $f_{\mathrm{BH}}$ for a sample with a mineral age of $t_{\mathrm{age}}=1.0\,\mathrm{Gyr}$. Note that this rate is a time-averaged value over the mineral age. We find that the required rate is lower when contributions from models of HNS formation and BH formation (failed SNe) are included. For the case of NS formation, since the total neutrino emission reflects the gravitational binding energy of the compact remnant, more massive remnants produce more neutrinos. Failed SNe are not included in the Galactic CC SN rate inferred from observed successful SNe; accordingly, they provide an additional contribution. As a result, the number of damage tracks in paleodetectors is larger for models with larger $f_{\mathrm{HNS}}$ and $f_{\mathrm{BH}}$ (see Fig.~\ref{fig:drdx}), leading to a stronger signal relative to the background and a lower required SN rate. Furthermore, as shown in Fig.~\ref{fig:drdx}, neutrinos from failed SNe produce more damage tracks at $x \gtrsim 100\,\mathrm{nm}$ than those from successful SNe, due to their higher average energies, thereby improving the sensitivity of paleodetectors.

In the cases of successful SNe, including both CNS and HNS formations, the required SN rate is the lowest for the Togashi EOS model and the highest for the Shen EOS model. Since the gravitational binding energy is larger for more compact NSs, EOSs with smaller NS radii produce more neutrinos. As a result, the Togashi EOS model leads to more damage tracks and a lower required SN rate, whereas the Shen EOS model yields fewer tracks and a higher required rate. Regarding the Shen EOS model, the minimum Galactic SN rate depends strongly on $f_{\mathrm{BH}}$. This behavior arises because the Shen EOS yields the largest neutrino emission among the three EOSs in the BH-forming (failed SN) case, owing to its large maximum proto-NS mass (see Sec.~\ref{sbsec:spectra}), while yielding the smallest emission in the NS-forming (successful SN) case.

The sensitivity as a function of $f_{\mathrm{HNS}}$ and $f_{\mathrm{BH}}$ is explored in Fig.~\ref{fig:R_min_contour}, where the minimum Galactic SN rate is shown as contour plots in the $f_{\mathrm{BH}}$--$f_{\mathrm{HNS}}$ plane. We find that the required Galactic SN rate decreases with increasing $f_{\mathrm{BH}}$ and $f_{\mathrm{HNS}}$, and that the dependence on these remnant fractions varies with the nuclear EOS. Nevertheless, for all EOSs considered,  the minimum detectable rate is reduced to $R_{\mathrm{SN}} \sim 1.0\times10^{-2}\,\mathrm{yr^{-1}}$ for the case of $(f_{\mathrm{BH}},f_{\mathrm{HNS}})=(0.2, 0.4)$, which is discussed in Sec.~\ref{sbsec:flux}.

In Fig.~\ref{fig:agemin}, we show the minimum mineral age required for a $3\sigma$ detection as a function of $f_{\mathrm{HNS}}$ and $f_{\mathrm{BH}}$, assuming a time-constant Galactic CC SN rate of $R_{\mathrm{SN}}=3.0\times10^{-2}\,\mathrm{yr^{-1}}$ \cite{2011MNRAS.412.1473L,2013ApJ...778..164A}. The required mineral age decreases with increasing $f_{\mathrm{HNS}}$ and $f_{\mathrm{BH}}$, as the enhanced neutrino emission in models including HNS and BH formation increases the number of damage tracks. The dependence on the nuclear EOS follows the same trend as seen in the minimum SN rate (see Fig.~\ref{fig:Rmin}). In successful SN cases, the Togashi EOS yields the largest neutrino emission due to its smaller NS radii, resulting in the shortest required mineral age. In contrast, for failed SNe, the Shen EOS leads to a significantly shorter required age depending on $f_{\mathrm{BH}}$, as it predicts the largest maximum proto-NS mass and hence the largest neutrino emission.

In Fig.~\ref{fig:agemin_contour}, the mineral age required for neutrino detection with paleodetectors is shown as a function of both $f_{\mathrm{HNS}}$ and $f_{\mathrm{BH}}$, assuming $R_{\mathrm{SN}}=3.0\times10^{-2}\,\mathrm{yr^{-1}}$. We find that the required age depends more strongly on $f_{\mathrm{BH}}$ than on $f_{\mathrm{HNS}}$ for all EOSs. For $(f_{\mathrm{BH}},f_{\mathrm{HNS}})=(0.2, 0.4)$, the required age is $t_{\mathrm{age}}<0.1\,\mathrm{Gyr}$, depending on the EOS.

\subsection{Burst-like CC SNe and snowball Earth event}\label{sbsec:snstv_burst}

\begin{figure*}
   \includegraphics[width=85mm]{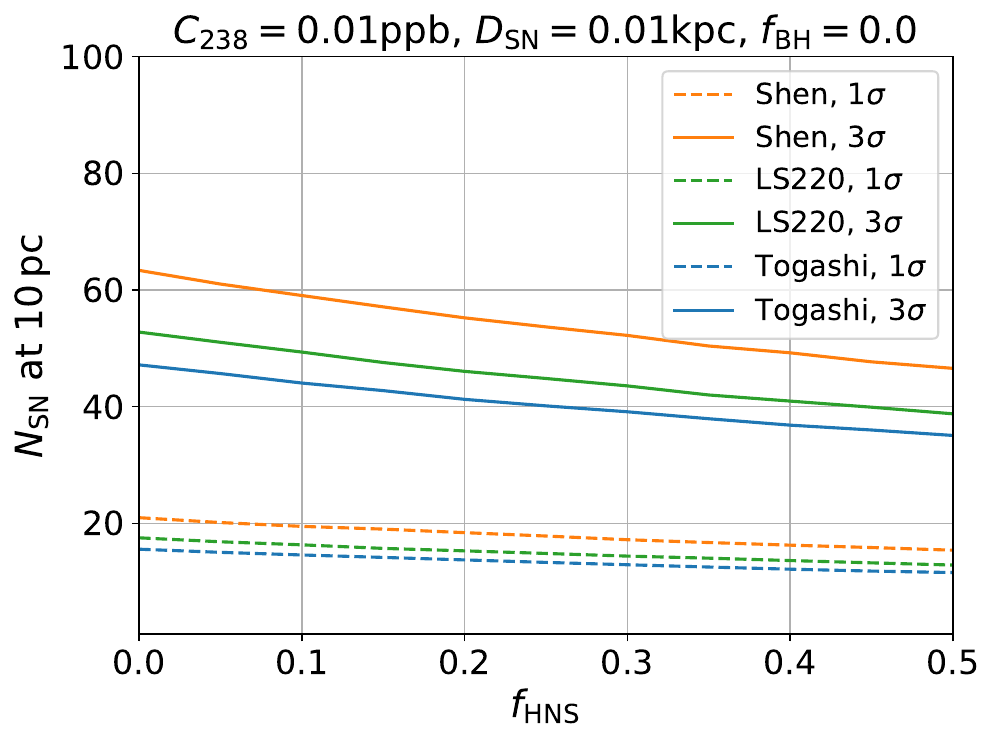}
   \includegraphics[width=85mm]{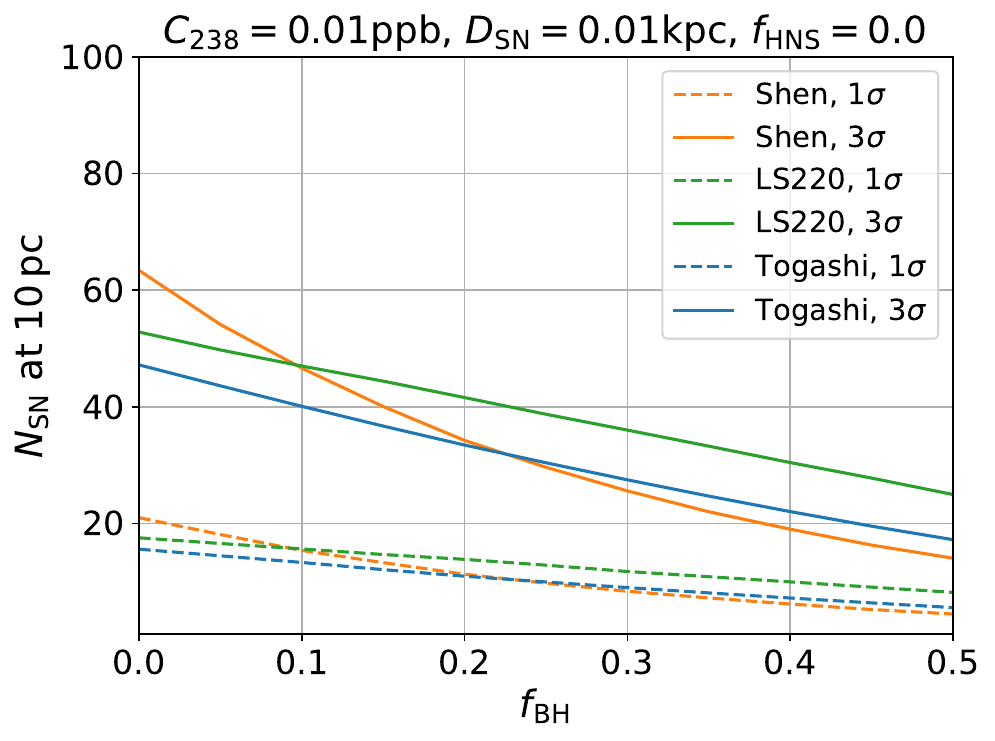}
   \caption{Left: Number of SNe required to detect burst-like events as a function of $f_{\mathrm{HNS}}$ with $f_{\mathrm{BH}}=0.0$. Right: Same as the left panel, but shown as a function of $f_{\mathrm{BH}}$ with $f_{\mathrm{HNS}}=0.0$. In both panels, the distance to the SNe is fixed to $D_{\mathrm{SN}}=10\,\mathrm{pc}$, and ten epsomite samples with ages $t_{\mathrm{age}}=0.1, 0.2, ..., 1.0\,\mathrm{Gyr}$, each with a mass of $M=100\,\mathrm{g}$ and a uranium-238 concentration of $C_{238}=0.01\,\mathrm{ppb}$, are assumed. Solid and dashed lines correspond to $3\sigma$ and $1\sigma$ detections, respectively. Line colors follow Fig.~\ref{fig:Rmin}.}
 \label{fig:N_SN_compare}
\end{figure*}

\begin{figure*}[htbp]
\centering

\begin{minipage}[t]{0.32\textwidth}
    \centering
    \includegraphics[width=\textwidth]{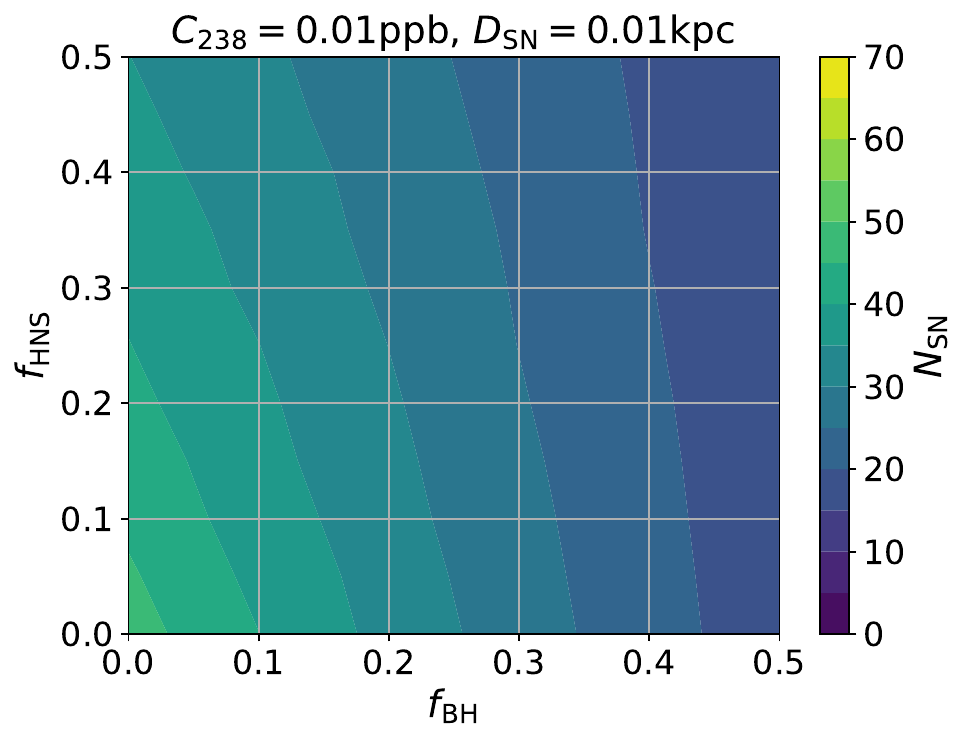}
\end{minipage}
\hfill
\begin{minipage}[t]{0.32\textwidth}
    \centering
    \includegraphics[width=\textwidth]{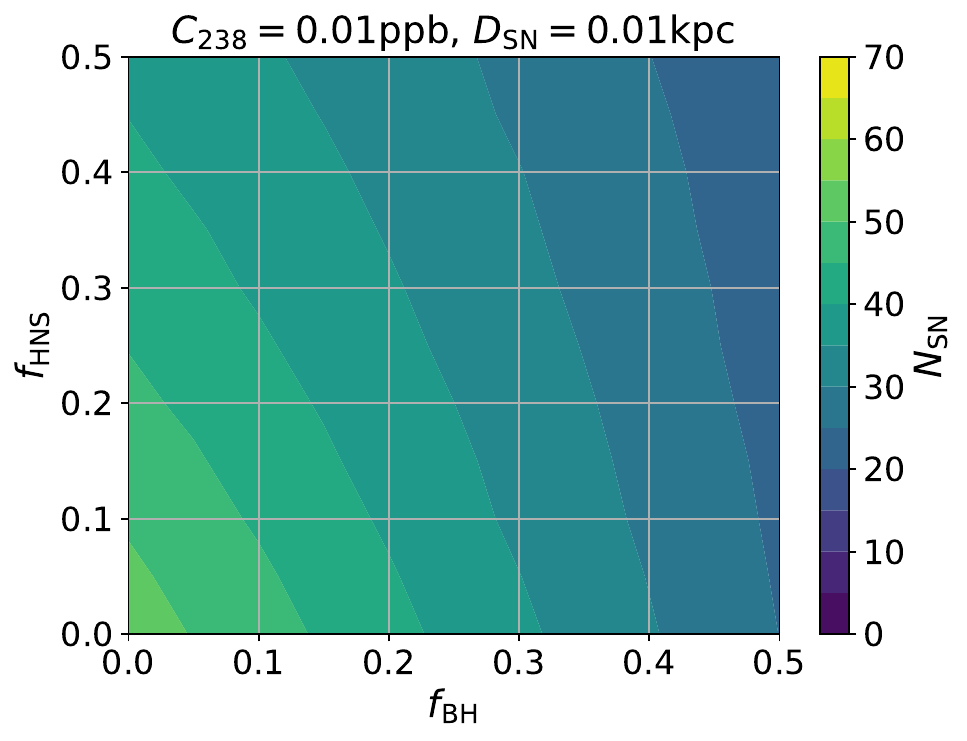}
\end{minipage}
\hfill
\begin{minipage}[t]{0.32\textwidth}
    \centering
    \includegraphics[width=\textwidth]{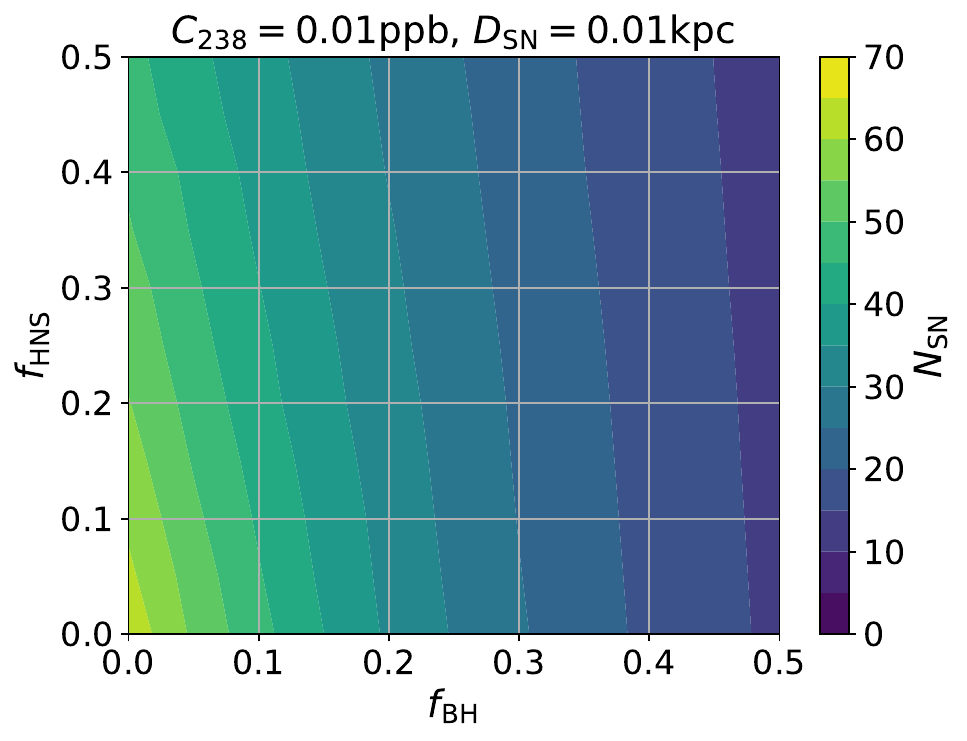}
\end{minipage}
    \caption{Number of SNe required to detect burst-like events as a function of $f_{\mathrm{BH}}$ and $f_{\mathrm{HNS}}$. The assumptions for the SN distance and the samples are the same as in Fig.~\ref{fig:N_SN_compare}. The left, center, and right panels show the results for the Togashi, LS220, and Shen EOS models, respectively.}
\label{fig:N_SN_contour}
\end{figure*}

In this subsection, we explore the sensitivity of paleodetectors to burst-like CC SN events that may be associated with a snowball Earth event. Following the experimental setup proposed in Ref.~\cite{2020PhRvD.101j3017B}, we assume ten epsomite samples with ages $t_{\mathrm{age}}=0.1, 0.2, ..., 1.0\,\mathrm{Gyr}$, each with a mass of $M=100\,\mathrm{g}$ and a uranium-238 concentration of $C_{238}=0.01\,\mathrm{ppb}$. We parametrize the burst-like CC SN events by their number, distance, and occurrence time, denoted by $N_{\mathrm{SN}}$, $D_{\mathrm{SN}}$, and $t_{\mathrm{SN}}$, respectively. Since the neutrino flux depends only on the combination $N_{\mathrm{SN}}/D_{\mathrm{SN}}^2$, we fix $D_{\mathrm{SN}}=10\,\mathrm{pc}$ as a benchmark and evaluate the sensitivity as a function of $N_{\mathrm{SN}}$. Under these assumptions, samples with $t_{\mathrm{age}} > t_{\mathrm{SN}}$ are expected to contain damage tracks from burst-like events in addition to the steady contribution associated with a time-constant CC SN rate. We generate mock data including the contribution of burst-like events and analyze them under the null hypothesis of a time-constant CC SN rate, in which the number of damage tracks scales with the mineral age. The sensitivity is quantified by the statistical significance with which this null hypothesis can be rejected.

We consider a snowball Earth event that occurred between $0.65\,\mathrm{Gyr}$ and $0.635\,\mathrm{Gyr}$ ago \cite{2019AREPS..47....1H}, and set $t_{\mathrm{SN}}=0.65\,\mathrm{Gyr}$. We also assume a time-constant CC SN rate of $R_{\mathrm{SN}}=3.0\times10^{-2}\,\mathrm{yr^{-1}}$. Under these assumptions, samples with ages $0.7$--$1.0\,\mathrm{Gyr}$ are expected to contain damage tracks from burst-like events in addition to the steady contribution.

In Fig.~\ref{fig:N_SN_compare}, we show the number of SNe required to detect burst-like events at the $1\sigma$ and $3\sigma$ levels for three nuclear EOSs. Including contributions from HNS-forming SNe and BH-forming failed SNe reduces the required $N_{\mathrm{SN}}$, as these components increase the neutrino emission (see Sec.~\ref{sbsec:snstv_neutrino}). We note that the diversity of core-collapse outcomes and the nuclear EOSs are taken into account for both the time-constant and burst-like scenarios simultaneously. Since the excess over the steady CC SN contribution becomes larger, the null hypothesis of a time-constant CC SN rate can be rejected more easily. The dependence on the nuclear EOS follows the same pattern as in Sec.~\ref{sbsec:snstv_neutrino} and can also be understood in terms of the total neutrino emission.

We evaluate the number of SNe required for a $3\sigma$ detection of burst-like events as a function of $f_{\mathrm{BH}}$ and $f_{\mathrm{HNS}}$ for three nuclear EOSs, and present the results as contour plots in Fig.~\ref{fig:N_SN_contour}. We find that the required $N_{\mathrm{SN}}$ is typically of order several tens across the parameter space. This suggests that, if a snowball Earth event were triggered by burst-like SN activity involving several tens of SNe at a distance of $10\,\mathrm{pc}$, such a scenario could be probed by paleodetectors. As discussed in Sec.~\ref{sec:intro}, this number is plausible in young massive clusters, which can host $N_{\mathrm{SN}} \gtrsim 100$ SNe.

\section{Conclusion}
\label{sec:conclusion}

Paleodetectors detect neutrinos emitted from Galactic core collapse supernovae (CC SNe) by utilizing nuclear recoil tracks preserved in mineral samples. Extending the analysis of Ref.~\cite{2020PhRvD.101j3017B}, we have investigated the sensitivity of paleodetectors to both the time-averaged Galactic CC SN rate and burst-like SN events. In particular, we have considered burst-like SN events associated with the snowball Earth epoch, motivated by the hypothesis that nearby SN activity may have played a role in triggering the snowball Earth events \cite{2020ApJ...904..137T}. To account for the diversity of stellar core-collapse outcomes, we have considered three types of remnants: canonical-mass neutron stars (NSs), high-mass NSs, and failed SNe leading to black hole (BH) formation. Following Ref.~\cite{2022ApJ...937...30A}, we have employed neutrino spectra derived from numerical simulations for three nuclear equation-of-state (EOS) models, and have parametrized the fractions of high-mass NSs and failed SNe by $f_{\mathrm{HNS}}$ and $f_{\mathrm{BH}}$. We have then evaluated the sensitivity as a function of these parameters and examined its dependence on the EOS.

In Sec.~\ref{sbsec:snstv_neutrino}, we have evaluated the minimum Galactic SN rate and the minimum mineral age required for neutrino detection. We have found that the sensitivity depends on the nuclear EOS and is enhanced in the case that contributions from high-mass NSs and failed SNe are included. This reflects the fact that the sensitivity increases with the neutrino emission. Among the three EOS models considered, the Togashi EOS yields the highest sensitivity for NS-forming cases due to its smaller NS radii, which correspond to larger neutrino emission. For failed SNe, the Shen EOS exhibits a strong dependence on $f_{\mathrm{BH}}$, reflecting its larger maximum proto-NS mass and enhanced neutrino emission.

In Sec.~\ref{sbsec:snstv_burst}, we have evaluated the sensitivity to burst-like SN events associated with the snowball Earth epoch. We have found that if several tens of SNe were to occur at a distance of $D_{\mathrm{SN}}=10\,\mathrm{pc}$ from Earth, such events would be detectable with paleodetectors, regardless of the stellar core-collapse diversity and nuclear EOS. Such a scenario may be realized in young massive clusters, which can host a large number of SNe within a few parsecs. Paleodetectors are therefore expected to provide a unique link between astrophysics and Earth science.

The data that support the findings of this article are openly available \cite{yamasaki_2026_21002077}.


\begin{acknowledgments}
This work was supported by Grants-in-Aid for Scientific Research (JP20K03973, JP24K07021) and Grant-in-Aid for Transformative Research Areas (JP24H02245) from the Ministry of Education, Culture, Sports, Science and Technology (MEXT), Japan.
\end{acknowledgments}



\bibliography{ref}

@ARTICLE{2020PhRvD.101j3017B,
       author = {{Baum}, Sebastian and {Edwards}, Thomas D.~P. and {Kavanagh}, Bradley J. and {Stengel}, Patrick and {Drukier}, Andrzej K. and {Freese}, Katherine and {G{\'o}rski}, Maciej and {Weniger}, Christoph},
        title = "{Paleodetectors for Galactic supernova neutrinos}",
      journal = {\prd},
         year = 2020,
        month = may,
       volume = {101},
       number = {10},
          eid = {103017},
        pages = {103017},
          doi = {10.1103/PhysRevD.101.103017},
archivePrefix = {arXiv},
       eprint = {1906.05800},
 primaryClass = {astro-ph.GA},
       adsurl = {https://ui.adsabs.harvard.edu/abs/2020PhRvD.101j3017B}
}

@ARTICLE{2003ApJ...590..971K,
       author = {{Keil}, Mathias Th. and {Raffelt}, Georg G. and {Janka}, Hans-Thomas},
        title = "{Monte Carlo Study of Supernova Neutrino Spectra Formation}",
      journal = {\apj},
         year = 2003,
        month = jun,
       volume = {590},
       number = {2},
        pages = {971-991},
          doi = {10.1086/375130},
archivePrefix = {arXiv},
       eprint = {astro-ph/0208035},
 primaryClass = {astro-ph},
       adsurl = {https://ui.adsabs.harvard.edu/abs/2003ApJ...590..971K}
}

@ARTICLE{2022ApJ...937...30A,
       author = {{Ashida}, Yosuke and {Nakazato}, Ken'ichiro},
        title = "{Exploring the Fate of Stellar Core Collapse with Supernova Relic Neutrinos}",
      journal = {\apj},
         year = 2022,
        month = sep,
       volume = {937},
       number = {1},
          eid = {30},
        pages = {30},
          doi = {10.3847/1538-4357/ac8a46},
archivePrefix = {arXiv},
       eprint = {2204.04880},
 primaryClass = {astro-ph.HE},
       adsurl = {https://ui.adsabs.harvard.edu/abs/2022ApJ...937...30A}
}

@ARTICLE{2018MNRAS.478.1377A,
       author = {{Alsing}, Justin and {Silva}, Hector O. and {Berti}, Emanuele},
        title = "{Evidence for a maximum mass cut-off in the neutron star mass distribution and constraints on the equation of state}",
      journal = {\mnras},
         year = 2018,
        month = jul,
       volume = {478},
       number = {1},
        pages = {1377-1391},
          doi = {10.1093/mnras/sty1065},
archivePrefix = {arXiv},
       eprint = {1709.07889},
 primaryClass = {astro-ph.HE},
       adsurl = {https://ui.adsabs.harvard.edu/abs/2018MNRAS.478.1377A}
}

@ARTICLE{2011ApJS..197...20S,
       author = {{Shen}, H. and {Toki}, H. and {Oyamatsu}, K. and {Sumiyoshi}, K.},
        title = "{Relativistic Equation of State for Core-collapse Supernova Simulations}",
      journal = {\apjs},
         year = 2011,
        month = dec,
       volume = {197},
       number = {2},
          eid = {20},
        pages = {20},
          doi = {10.1088/0067-0049/197/2/20},
archivePrefix = {arXiv},
       eprint = {1105.1666},
 primaryClass = {astro-ph.HE},
       adsurl = {https://ui.adsabs.harvard.edu/abs/2011ApJS..197...20S}
}

@ARTICLE{1991NuPhA.535..331L,
       author = {{Lattimer}, James M. and {Swesty}, Douglas F.},
        title = "{A generalized equation of state for hot, dense matter}",
      journal = {\nphysa},
         year = 1991,
        month = dec,
       volume = {535},
       number = {2},
        pages = {331-376},
          doi = {10.1016/0375-9474(91)90452-C},
       adsurl = {https://ui.adsabs.harvard.edu/abs/1991NuPhA.535..331L}
}

@ARTICLE{2017NuPhA.961...78T,
       author = {{Togashi}, H. and {Nakazato}, K. and {Takehara}, Y. and {Yamamuro}, S. and {Suzuki}, H. and {Takano}, M.},
        title = "{Nuclear equation of state for core-collapse supernova simulations with realistic nuclear forces}",
      journal = {\nphysa},
         year = 2017,
        month = may,
       volume = {961},
        pages = {78-105},
          doi = {10.1016/j.nuclphysa.2017.02.010},
archivePrefix = {arXiv},
       eprint = {1702.05324},
 primaryClass = {nucl-th},
       adsurl = {https://ui.adsabs.harvard.edu/abs/2017NuPhA.961...78T}
}

@ARTICLE{2019AREPS..47....1H,
       author = {{Hoffman}, Paul F.},
        title = "{Big Time}",
      journal = {Annual Review of Earth and Planetary Sciences},
         year = 2019,
        month = may,
       volume = {47},
        pages = {1-17},
          doi = {10.1146/annurev-earth-053018-060145},
       adsurl = {https://ui.adsabs.harvard.edu/abs/2019AREPS..47....1H}
}

@ARTICLE{2010ARA&A..48..431P,
       author = {{Portegies Zwart}, Simon F. and {McMillan}, Stephen L.~W. and {Gieles}, Mark},
        title = "{Young Massive Star Clusters}",
      journal = {\araa},
         year = 2010,
        month = sep,
       volume = {48},
        pages = {431-493},
          doi = {10.1146/annurev-astro-081309-130834},
archivePrefix = {arXiv},
       eprint = {1002.1961},
 primaryClass = {astro-ph.GA},
       adsurl = {https://ui.adsabs.harvard.edu/abs/2010ARA&A..48..431P}
}

@ARTICLE{2014ARA&A..52..415M,
       author = {{Madau}, Piero and {Dickinson}, Mark},
        title = "{Cosmic Star-Formation History}",
      journal = {\araa},
         year = 2014,
        month = aug,
       volume = {52},
        pages = {415-486},
          doi = {10.1146/annurev-astro-081811-125615},
archivePrefix = {arXiv},
       eprint = {1403.0007},
 primaryClass = {astro-ph.CO},
       adsurl = {https://ui.adsabs.harvard.edu/abs/2014ARA&A..52..415M}
}

@ARTICLE{2020ApJ...904..137T,
       author = {{Tsujimoto}, Takuji and {Baba}, Junichi},
        title = "{Remarkable Migration of the Solar System from the Innermost Galactic Disk; a Wander, a Wobble, and a Climate Catastrophe on the Earth}",
      journal = {\apj},
         year = 2020,
        month = dec,
       volume = {904},
       number = {2},
          eid = {137},
        pages = {137},
          doi = {10.3847/1538-4357/abc00a},
archivePrefix = {arXiv},
       eprint = {2010.05962},
 primaryClass = {astro-ph.GA},
       adsurl = {https://ui.adsabs.harvard.edu/abs/2020ApJ...904..137T}
}

@ARTICLE{2023PDU....4101245B,
       author = {{Baum}, Sebastian and {Stengel}, Patrick and {Abe}, Natsue and {Acevedo}, Javier F. and {Araujo}, Gabriela R. and {Asahara}, Yoshihiro and {Avignone}, Frank and {Balogh}, Levente and {Baudis}, Laura and {Boukhtouchen}, Yilda and {Bramante}, Joseph and {Breur}, Pieter Alexander and {Caccianiga}, Lorenzo and {Capozzi}, Francesco and {Collar}, Juan I. and {Ebadi}, Reza and {Edwards}, Thomas and {Eitel}, Klaus and {Elykov}, Alexey and {Ewing}, Rodney C. and {Freese}, Katherine and {Fung}, Audrey and {Galelli}, Claudio and {Glasmacher}, Ulrich A. and {Gleason}, Arianna and {Hasebe}, Noriko and {Hirose}, Shigenobu and {Horiuchi}, Shunsaku and {Hoshino}, Yasushi and {Huber}, Patrick and {Ido}, Yuki and {Igami}, Yohei and {Ishikawa}, Norito and {Itow}, Yoshitaka and {Kamiyama}, Takashi and {Kato}, Takenori and {Kavanagh}, Bradley J. and {Kawamura}, Yoji and {Kazama}, Shingo and {Kenney}, Christopher J. and {Kilminster}, Ben and {Kouketsu}, Yui and {Kozaka}, Yukiko and {Kurinsky}, Noah A. and {Leybourne}, Matthew and {Lucas}, Thalles and {McDonough}, William F. and {Marshall}, Mason C. and {Mateos}, Jose Maria and {Mathur}, Anubhav and {Michibayashi}, Katsuyoshi and {Mkhonto}, Sharlotte and {Murase}, Kohta and {Naka}, Tatsuhiro and {Oguni}, Kenji and {Rajendran}, Surjeet and {Sakane}, Hitoshi and {Sala}, Paola and {Scholberg}, Kate and {Semenec}, Ingrida and {Shiraishi}, Takuya and {Spitz}, Joshua and {Sun}, Kai and {Suzuki}, Katsuhiko and {Tanin}, Erwin H. and {Vincent}, Aaron and {Vladimirov}, Nikita and {Walsworth}, Ronald L. and {Watanabe}, Hiroko},
        title = "{Mineral detection of neutrinos and dark matter. A whitepaper}",
      journal = {Physics of the Dark Universe},
         year = 2023,
        month = aug,
       volume = {41},
          eid = {101245},
        pages = {101245},
          doi = {10.1016/j.dark.2023.101245},
archivePrefix = {arXiv},
       eprint = {2301.07118},
 primaryClass = {astro-ph.IM},
       adsurl = {https://ui.adsabs.harvard.edu/abs/2023PDU....4101245B}
}

@ARTICLE{2011MNRAS.412.1473L,
       author = {{Li}, Weidong and {Chornock}, Ryan and {Leaman}, Jesse and {Filippenko}, Alexei V. and {Poznanski}, Dovi and {Wang}, Xiaofeng and {Ganeshalingam}, Mohan and {Mannucci}, Filippo},
        title = "{Nearby supernova rates from the Lick Observatory Supernova Search - III. The rate-size relation, and the rates as a function of galaxy Hubble type and colour}",
      journal = {\mnras},
         year = 2011,
        month = apr,
       volume = {412},
       number = {3},
        pages = {1473-1507},
          doi = {10.1111/j.1365-2966.2011.18162.x},
archivePrefix = {arXiv},
       eprint = {1006.4613},
 primaryClass = {astro-ph.SR},
       adsurl = {https://ui.adsabs.harvard.edu/abs/2011MNRAS.412.1473L}
}

@ARTICLE{2013ApJ...778..164A,
       author = {{Adams}, Scott M. and {Kochanek}, C.~S. and {Beacom}, John F. and {Vagins}, Mark R. and {Stanek}, K.~Z.},
        title = "{Observing the Next Galactic Supernova}",
      journal = {\apj},
         year = 2013,
        month = dec,
       volume = {778},
       number = {2},
          eid = {164},
        pages = {164},
          doi = {10.1088/0004-637X/778/2/164},
archivePrefix = {arXiv},
       eprint = {1306.0559},
 primaryClass = {astro-ph.HE},
       adsurl = {https://ui.adsabs.harvard.edu/abs/2013ApJ...778..164A}
}

@ARTICLE{2021NewA...8301498R,
       author = {{Rozwadowska}, Karolina and {Vissani}, Francesco and {Cappellaro}, Enrico},
        title = "{On the rate of core collapse supernovae in the milky way}",
      journal = {\na},
         year = 2021,
        month = feb,
       volume = {83},
          eid = {101498},
        pages = {101498},
          doi = {10.1016/j.newast.2020.101498},
archivePrefix = {arXiv},
       eprint = {2009.03438},
 primaryClass = {astro-ph.HE},
       adsurl = {https://ui.adsabs.harvard.edu/abs/2021NewA...8301498R}
}

@ARTICLE{2025MNRAS.538.1367Q,
       author = {{Quintana}, Alexis L. and {Wright}, Nicholas J. and {Mart{\'\i}nez Garc{\'\i}a}, Juan},
        title = "{A census of OB stars within 1 kpc and the star formation and core collapse supernova rates of the Milky Way}",
      journal = {\mnras},
         year = 2025,
        month = apr,
       volume = {538},
       number = {3},
        pages = {1367-1383},
          doi = {10.1093/mnras/staf083},
archivePrefix = {arXiv},
       eprint = {2503.08286},
 primaryClass = {astro-ph.SR},
       adsurl = {https://ui.adsabs.harvard.edu/abs/2025MNRAS.538.1367Q}
}

@ARTICLE{2002RvMP...74.1015W,
       author = {{Woosley}, S.~E. and {Heger}, A. and {Weaver}, T.~A.},
        title = "{The evolution and explosion of massive stars}",
      journal = {Reviews of Modern Physics},
         year = 2002,
        month = nov,
       volume = {74},
       number = {4},
        pages = {1015-1071},
          doi = {10.1103/RevModPhys.74.1015},
       adsurl = {https://ui.adsabs.harvard.edu/abs/2002RvMP...74.1015W}
}

@ARTICLE{2009MNRAS.395.1409S,
       author = {{Smartt}, S.~J. and {Eldridge}, J.~J. and {Crockett}, R.~M. and {Maund}, J.~R.},
        title = "{The death of massive stars - I. Observational constraints on the progenitors of Type II-P supernovae}",
      journal = {\mnras},
         year = 2009,
        month = may,
       volume = {395},
       number = {3},
        pages = {1409-1437},
          doi = {10.1111/j.1365-2966.2009.14506.x},
archivePrefix = {arXiv},
       eprint = {0809.0403},
 primaryClass = {astro-ph},
       adsurl = {https://ui.adsabs.harvard.edu/abs/2009MNRAS.395.1409S}
}

@ARTICLE{2017A&A...601A..29Z,
       author = {{Zapartas}, E. and {de Mink}, S.~E. and {Izzard}, R.~G. and {Yoon}, S.-C. and {Badenes}, C. and {G{\"o}tberg}, Y. and {de Koter}, A. and {Neijssel}, C.~J. and {Renzo}, M. and {Schootemeijer}, A. and {Shrotriya}, T.~S.},
        title = "{Delay-time distribution of core-collapse supernovae with late events resulting from binary interaction}",
      journal = {\aap},
         year = 2017,
        month = may,
       volume = {601},
          eid = {A29},
        pages = {A29},
          doi = {10.1051/0004-6361/201629685},
archivePrefix = {arXiv},
       eprint = {1701.07032},
 primaryClass = {astro-ph.HE},
       adsurl = {https://ui.adsabs.harvard.edu/abs/2017A&A...601A..29Z}
}

@ARTICLE{2019ARA&A..57..227K,
       author = {{Krumholz}, Mark R. and {McKee}, Christopher F. and {Bland-Hawthorn}, Joss},
        title = "{Star Clusters Across Cosmic Time}",
      journal = {\araa},
         year = 2019,
        month = aug,
       volume = {57},
        pages = {227-303},
          doi = {10.1146/annurev-astro-091918-104430},
archivePrefix = {arXiv},
       eprint = {1812.01615},
 primaryClass = {astro-ph.GA},
       adsurl = {https://ui.adsabs.harvard.edu/abs/2019ARA&A..57..227K}
}

@ARTICLE{1996A&A...314..438W,
       author = {{Wielen}, R. and {Fuchs}, B. and {Dettbarn}, C.},
        title = "{On the birth-place of the Sun and the places of formation of other nearby stars}",
      journal = {\aap},
         year = 1996,
        month = oct,
       volume = {314},
        pages = {438},
       adsurl = {https://ui.adsabs.harvard.edu/abs/1996A&A...314..438W}
}

@ARTICLE{2012A&A...539A.143N,
       author = {{Nieva}, M.-F. and {Przybilla}, N.},
        title = "{Present-day cosmic abundances. A comprehensive study of nearby early B-type stars and implications for stellar and Galactic evolution and interstellar dust models}",
      journal = {\aap},
         year = 2012,
        month = mar,
       volume = {539},
          eid = {A143},
        pages = {A143},
          doi = {10.1051/0004-6361/201118158},
archivePrefix = {arXiv},
       eprint = {1203.5787},
 primaryClass = {astro-ph.SR},
       adsurl = {https://ui.adsabs.harvard.edu/abs/2012A&A...539A.143N}
}

@ARTICLE{2024ApJ...976L..29B,
       author = {{Baba}, Junichi and {Tsujimoto}, Takuji and {Saitoh}, Takayuki R.},
        title = "{Solar System Migration Points to a Renewed Concept: Galactic Habitable Orbits}",
      journal = {\apjl},
         year = 2024,
        month = dec,
       volume = {976},
       number = {2},
          eid = {L29},
        pages = {L29},
          doi = {10.3847/2041-8213/ad9260},
archivePrefix = {arXiv},
       eprint = {2412.02963},
 primaryClass = {astro-ph.GA},
       adsurl = {https://ui.adsabs.harvard.edu/abs/2024ApJ...976L..29B}
}

@ARTICLE{2002PhRvL..89e1102S,
       author = {{Shaviv}, Nir J.},
        title = "{Cosmic Ray Diffusion from the Galactic Spiral Arms, Iron Meteorites, and a Possible Climatic Connection}",
      journal = {\prl},
         year = 2002,
        month = jan,
       volume = {89},
       number = {5},
          eid = {051102},
        pages = {051102},
          doi = {10.1103/PhysRevLett.89.051102},
archivePrefix = {arXiv},
       eprint = {astro-ph/0207637},
 primaryClass = {astro-ph},
       adsurl = {https://ui.adsabs.harvard.edu/abs/2002PhRvL..89e1102S}
}

@ARTICLE{2007A&G....48a..18S,
       author = {{Svensmark}, Henrik},
        title = "{Cosmoclimatology: a new theory emerges}",
      journal = {Astronomy and Geophysics},
         year = 2007,
        month = feb,
       volume = {48},
       number = {1},
        pages = {1.18-1.24},
          doi = {10.1111/j.1468-4004.2007.48118.x},
       adsurl = {https://ui.adsabs.harvard.edu/abs/2007A&G....48a..18S}
}

@ARTICLE{2017NatCo...8.2199S,
       author = {{Svensmark}, H. and {Enghoff}, M.~B. and {Shaviv}, N.~J. and {Svensmark}, J.},
        title = "{Increased ionization supports growth of aerosols into cloud condensation nuclei}",
      journal = {Nature Communications},
         year = 2017,
        month = dec,
       volume = {8},
          eid = {2199},
        pages = {2199},
          doi = {10.1038/s41467-017-02082-2},
       adsurl = {https://ui.adsabs.harvard.edu/abs/2017NatCo...8.2199S}
}

@ARTICLE{2021ApJ...909..169K,
       author = {{Kresse}, Daniel and {Ertl}, Thomas and {Janka}, Hans-Thomas},
        title = "{Stellar Collapse Diversity and the Diffuse Supernova Neutrino Background}",
      journal = {\apj},
         year = 2021,
        month = mar,
       volume = {909},
       number = {2},
          eid = {169},
        pages = {169},
          doi = {10.3847/1538-4357/abd54e},
archivePrefix = {arXiv},
       eprint = {2010.04728},
 primaryClass = {astro-ph.HE},
       adsurl = {https://ui.adsabs.harvard.edu/abs/2021ApJ...909..169K}
}

@ARTICLE{2023ApJ...953..151A,
       author = {{Ashida}, Yosuke and {Nakazato}, Ken'ichiro and {Tsujimoto}, Takuji},
        title = "{Diffuse Neutrino Flux Based on the Rates of Core-collapse Supernovae and Black Hole Formation Deduced from a Novel Galactic Chemical Evolution Model}",
      journal = {\apj},
         year = 2023,
        month = aug,
       volume = {953},
       number = {2},
          eid = {151},
        pages = {151},
          doi = {10.3847/1538-4357/ace3ba},
archivePrefix = {arXiv},
       eprint = {2305.13543},
 primaryClass = {astro-ph.HE},
       adsurl = {https://ui.adsabs.harvard.edu/abs/2023ApJ...953..151A}
}

@ARTICLE{2004ApJS..150..263L,
       author = {{Liebend{\"o}rfer}, Matthias and {Messer}, O.~E. Bronson and {Mezzacappa}, Anthony and {Bruenn}, Stephen W. and {Cardall}, Christian Y. and {Thielemann}, F.-K.},
        title = "{A Finite Difference Representation of Neutrino Radiation Hydrodynamics in Spherically Symmetric General Relativistic Spacetime}",
      journal = {\apjs},
         year = 2004,
        month = jan,
       volume = {150},
       number = {1},
        pages = {263-316},
          doi = {10.1086/380191},
archivePrefix = {arXiv},
       eprint = {astro-ph/0207036},
 primaryClass = {astro-ph},
       adsurl = {https://ui.adsabs.harvard.edu/abs/2004ApJS..150..263L}
}

@ARTICLE{2006PhRvL..97i1101S,
       author = {{Sumiyoshi}, K. and {Yamada}, S. and {Suzuki}, H. and {Chiba}, S.},
        title = "{Neutrino Signals from the Formation of a Black Hole: A Probe of the Equation of State of Dense Matter}",
      journal = {\prl},
         year = 2006,
        month = sep,
       volume = {97},
       number = {9},
          eid = {091101},
        pages = {091101},
          doi = {10.1103/PhysRevLett.97.091101},
archivePrefix = {arXiv},
       eprint = {astro-ph/0608509},
 primaryClass = {astro-ph},
       adsurl = {https://ui.adsabs.harvard.edu/abs/2006PhRvL..97i1101S}
}

@ARTICLE{2020PhRvD.101l3013W,
       author = {{Walk}, Laurie and {Tamborra}, Irene and {Janka}, Hans-Thomas and {Summa}, Alexander and {Kresse}, Daniel},
        title = "{Neutrino emission characteristics of black hole formation in three-dimensional simulations of stellar collapse}",
      journal = {\prd},
         year = 2020,
        month = jun,
       volume = {101},
       number = {12},
          eid = {123013},
        pages = {123013},
          doi = {10.1103/PhysRevD.101.123013},
archivePrefix = {arXiv},
       eprint = {1910.12971},
 primaryClass = {astro-ph.HE},
       adsurl = {https://ui.adsabs.harvard.edu/abs/2020PhRvD.101l3013W}
}

@ARTICLE{2023Univ...10....3R,
       author = {{Rocha}, L{\'\i}via S. and {Horvath}, Jorge E. and {de S{\'a}}, Lucas M. and {Chinen}, Gustavo Y. and {Bar{\~a}o}, Lucas G. and {de Avellar}, Marcio G.~B.},
        title = "{Mass Distribution and Maximum Mass of Neutron Stars: Effects of Orbital Inclination Angle}",
      journal = {Universe},
         year = 2023,
        month = dec,
       volume = {10},
       number = {1},
          eid = {3},
        pages = {3},
          doi = {10.3390/universe10010003},
archivePrefix = {arXiv},
       eprint = {2312.13244},
 primaryClass = {astro-ph.HE},
       adsurl = {https://ui.adsabs.harvard.edu/abs/2023Univ...10....3R}
}

@ARTICLE{2024PhRvD.109d3052F,
       author = {{Fan}, Yi-Zhong and {Han}, Ming-Zhe and {Jiang}, Jin-Liang and {Shao}, Dong-Sheng and {Tang}, Shao-Peng},
        title = "{Maximum gravitational mass M$_{TOV}$=2.2 5$_{-0.07}$$^{+0.08}$M$_{{\ensuremath{\odot}}}$ inferred at about 3\% precision with multimessenger data of neutron stars}",
      journal = {\prd},
         year = 2024,
        month = feb,
       volume = {109},
       number = {4},
          eid = {043052},
        pages = {043052},
          doi = {10.1103/PhysRevD.109.043052},
archivePrefix = {arXiv},
       eprint = {2309.12644},
 primaryClass = {astro-ph.HE},
       adsurl = {https://ui.adsabs.harvard.edu/abs/2024PhRvD.109d3052F}
}

@ARTICLE{2025PhRvD.111b3029G,
       author = {{Golomb}, Jacob and {Legred}, Isaac and {Chatziioannou}, Katerina and {Landry}, Philippe},
        title = "{Interplay of astrophysics and nuclear physics in determining the properties of neutron stars}",
      journal = {\prd},
         year = 2025,
        month = jan,
       volume = {111},
       number = {2},
          eid = {023029},
        pages = {023029},
          doi = {10.1103/PhysRevD.111.023029},
archivePrefix = {arXiv},
       eprint = {2410.14597},
 primaryClass = {astro-ph.HE},
       adsurl = {https://ui.adsabs.harvard.edu/abs/2025PhRvD.111b3029G}
}

@ARTICLE{2025PhRvD.112b3045B,
       author = {{Biswas}, Bhaskar and {Rosswog}, Stephan},
        title = "{Simultaneously constraining the neutron star equation of state and mass distribution through multimessenger observations and nuclear benchmarks}",
      journal = {\prd},
         year = 2025,
        month = jul,
       volume = {112},
       number = {2},
          eid = {023045},
        pages = {023045},
          doi = {10.1103/8lv3-1ywb},
archivePrefix = {arXiv},
       eprint = {2408.15192},
 primaryClass = {astro-ph.HE},
       adsurl = {https://ui.adsabs.harvard.edu/abs/2025PhRvD.112b3045B}
}

@ARTICLE{2016ARA&A..54..401O,
       author = {{{\"O}zel}, Feryal and {Freire}, Paulo},
        title = "{Masses, Radii, and the Equation of State of Neutron Stars}",
      journal = {\araa},
         year = 2016,
        month = sep,
       volume = {54},
        pages = {401-440},
          doi = {10.1146/annurev-astro-081915-023322},
archivePrefix = {arXiv},
       eprint = {1603.02698},
 primaryClass = {astro-ph.HE},
       adsurl = {https://ui.adsabs.harvard.edu/abs/2016ARA&A..54..401O}
}

@ARTICLE{2021ARNPS..71..433L,
       author = {{Lattimer}, J.~M.},
        title = "{Neutron Stars and the Nuclear Matter Equation of State}",
      journal = {Annual Review of Nuclear and Particle Science},
         year = 2021,
        month = sep,
       volume = {71},
        pages = {433-464},
          doi = {10.1146/annurev-nucl-102419-124827},
       adsurl = {https://ui.adsabs.harvard.edu/abs/2021ARNPS..71..433L}
}

@ARTICLE{2024PhRvD.110j3040L,
       author = {{Li}, Bao-An and {Grundler}, Xavier and {Xie}, Wen-Jie and {Zhang}, Nai-Bo},
        title = "{Bayesian inference of fine features of the nuclear equation of state from future neutron star radius measurements to 0.1 km accuracy}",
      journal = {\prd},
         year = 2024,
        month = nov,
       volume = {110},
       number = {10},
          eid = {103040},
        pages = {103040},
          doi = {10.1103/PhysRevD.110.103040},
archivePrefix = {arXiv},
       eprint = {2407.07823},
 primaryClass = {astro-ph.HE},
       adsurl = {https://ui.adsabs.harvard.edu/abs/2024PhRvD.110j3040L}
}

@ARTICLE{2025PhRvX..15b1014K,
       author = {{Koehn}, Hauke and {Rose}, Henrik and {Pang}, Peter T.~H. and {Somasundaram}, Rahul and {Reed}, Brendan T. and {Tews}, Ingo and {Abac}, Adrian and {Komoltsev}, Oleg and {Kunert}, Nina and {Kurkela}, Aleksi and {Coughlin}, Michael W. and {Healy}, Brian F. and {Dietrich}, Tim},
        title = "{From Existing and New Nuclear and Astrophysical Constraints to Stringent Limits on the Equation of State of Neutron-Rich Dense Matter}",
      journal = {Physical Review X},
         year = 2025,
        month = apr,
       volume = {15},
       number = {2},
          eid = {021014},
        pages = {021014},
          doi = {10.1103/PhysRevX.15.021014},
archivePrefix = {arXiv},
       eprint = {2402.04172},
 primaryClass = {astro-ph.HE},
       adsurl = {https://ui.adsabs.harvard.edu/abs/2025PhRvX..15b1014K}
}

@ARTICLE{2007ApJ...667..382S,
       author = {{Sumiyoshi}, K. and {Yamada}, S. and {Suzuki}, H.},
        title = "{Dynamics and Neutrino Signal of Black Hole Formation in Nonrotating Failed Supernovae. I. Equation of State Dependence}",
      journal = {\apj},
         year = 2007,
        month = sep,
       volume = {667},
       number = {1},
        pages = {382-394},
          doi = {10.1086/520876},
archivePrefix = {arXiv},
       eprint = {0706.3762},
 primaryClass = {astro-ph},
       adsurl = {https://ui.adsabs.harvard.edu/abs/2007ApJ...667..382S}
}

@ARTICLE{2013ApJ...774...17S,
       author = {{Steiner}, A.~W. and {Hempel}, M. and {Fischer}, T.},
        title = "{Core-collapse Supernova Equations of State Based on Neutron Star Observations}",
      journal = {\apj},
         year = 2013,
        month = sep,
       volume = {774},
       number = {1},
          eid = {17},
        pages = {17},
          doi = {10.1088/0004-637X/774/1/17},
archivePrefix = {arXiv},
       eprint = {1207.2184},
 primaryClass = {astro-ph.SR},
       adsurl = {https://ui.adsabs.harvard.edu/abs/2013ApJ...774...17S}
}

@ARTICLE{2020ApJ...894....4D,
       author = {{da Silva Schneider}, Andr{\'e} and {O'Connor}, Evan and {Granqvist}, Elvira and {Betranhandy}, Aurore and {Couch}, Sean M.},
        title = "{Equation of State and Progenitor Dependence of Stellar-mass Black Hole Formation}",
      journal = {\apj},
         year = 2020,
        month = may,
       volume = {894},
       number = {1},
          eid = {4},
        pages = {4},
          doi = {10.3847/1538-4357/ab8308},
archivePrefix = {arXiv},
       eprint = {2001.10434},
 primaryClass = {astro-ph.HE},
       adsurl = {https://ui.adsabs.harvard.edu/abs/2020ApJ...894....4D}
}

@ARTICLE{2025OJAp....8E.167S,
       author = {{Suwa}, Yudai and {Akaho}, Ryuichiro and {Ashida}, Yosuke and {Harada}, Akira and {Harada}, Masayuki and {Koshio}, Yusuke and {Mori}, Masamitsu and {Nakanishi}, Fumi and {Nakazato}, Ken'ichiro and {Sumiyoshi}, Kohsuke and {Wendell}, Roger A. and {Zaizen}, Masamichi},
        title = "{Neutrino Constraints on Black Hole Formation in M31}",
      journal = {The Open Journal of Astrophysics},
         year = 2025,
        month = nov,
       volume = {8},
        pages = {E167},
          doi = {10.33232/001c.147127},
archivePrefix = {arXiv},
       eprint = {2504.19510},
 primaryClass = {astro-ph.HE},
       adsurl = {https://ui.adsabs.harvard.edu/abs/2025OJAp....8E.167S}
}

@ARTICLE{2020ApJ...896...56W,
       author = {{Woosley}, S.~E. and {Sukhbold}, Tuguldur and {Janka}, H.-T.},
        title = "{The Birth Function for Black Holes and Neutron Stars in Close Binaries}",
      journal = {\apj},
         year = 2020,
        month = jun,
       volume = {896},
       number = {1},
          eid = {56},
        pages = {56},
          doi = {10.3847/1538-4357/ab8cc1},
archivePrefix = {arXiv},
       eprint = {2001.10492},
 primaryClass = {astro-ph.HE},
       adsurl = {https://ui.adsabs.harvard.edu/abs/2020ApJ...896...56W}
}

@ARTICLE{2021MNRAS.508..516N,
       author = {{Neustadt}, J.~M.~M. and {Kochanek}, C.~S. and {Stanek}, K.~Z. and {Basinger}, C. and {Jayasinghe}, T. and {Garling}, C.~T. and {Adams}, S.~M. and {Gerke}, J.},
        title = "{The search for failed supernovae with the Large Binocular Telescope: a new candidate and the failed SN fraction with 11 yr of data}",
      journal = {\mnras},
         year = 2021,
        month = nov,
       volume = {508},
       number = {1},
        pages = {516-528},
          doi = {10.1093/mnras/stab2605},
archivePrefix = {arXiv},
       eprint = {2104.03318},
 primaryClass = {astro-ph.SR},
       adsurl = {https://ui.adsabs.harvard.edu/abs/2021MNRAS.508..516N}
}

@software{thomas_edwards_2019_3245799,
  author       = {Thomas Edwards and
                  Bradley J. Kavanagh},
  title        = {tedwards2412/SN-paleology: Version 1.0 - ArXiv
                   release
                  },
  month        = jun,
  year         = 2019,
  publisher    = {Zenodo},
  version      = {1.0},
  doi          = {10.5281/zenodo.3245799},
  url          = {https://doi.org/10.5281/zenodo.3245799},
}

@ARTICLE{2017arXiv171205401E,
       author = {{Edwards}, Thomas D.~P. and {Weniger}, Christoph},
        title = "{swordfish: Efficient Forecasting of New Physics Searches without Monte Carlo}",
      journal = {arXiv e-prints},
         year = 2017,
        month = dec,
          eid = {arXiv:1712.05401},
        pages = {arXiv:1712.05401},
          doi = {10.48550/arXiv.1712.05401},
archivePrefix = {arXiv},
       eprint = {1712.05401},
 primaryClass = {hep-ph},
       adsurl = {https://ui.adsabs.harvard.edu/abs/2017arXiv171205401E}
}

@ARTICLE{2018JCAP...02..021E,
       author = {{Edwards}, Thomas D.~P. and {Weniger}, Christoph},
        title = "{A fresh approach to forecasting in astroparticle physics and dark matter searches}",
      journal = {\jcap},
         year = 2018,
        month = feb,
       volume = {2018},
       number = {2},
          eid = {021},
        pages = {021},
          doi = {10.1088/1475-7516/2018/02/021},
archivePrefix = {arXiv},
       eprint = {1704.05458},
 primaryClass = {astro-ph.IM},
       adsurl = {https://ui.adsabs.harvard.edu/abs/2018JCAP...02..021E}
}

@ARTICLE{2019PhRvD..99d3014D,
       author = {{Drukier}, Andrzej K. and {Baum}, Sebastian and {Freese}, Katherine and {G{\'o}rski}, Maciej and {Stengel}, Patrick},
        title = "{Paleo-detectors: Searching for dark matter with ancient minerals}",
      journal = {\prd},
         year = 2019,
        month = feb,
       volume = {99},
       number = {4},
          eid = {043014},
        pages = {043014},
          doi = {10.1103/PhysRevD.99.043014},
archivePrefix = {arXiv},
       eprint = {1811.06844},
 primaryClass = {astro-ph.CO},
       adsurl = {https://ui.adsabs.harvard.edu/abs/2019PhRvD..99d3014D}
}

@ARTICLE{2019PhRvD..99d3541E,
       author = {{Edwards}, Thomas D.~P. and {Kavanagh}, Bradley J. and {Weniger}, Christoph and {Baum}, Sebastian and {Drukier}, Andrzej K. and {Freese}, Katherine and {G{\'o}rski}, Maciej and {Stengel}, Patrick},
        title = "{Digging for dark matter: Spectral analysis and discovery potential of paleo-detectors}",
      journal = {\prd},
         year = 2019,
        month = feb,
       volume = {99},
       number = {4},
          eid = {043541},
        pages = {043541},
          doi = {10.1103/PhysRevD.99.043541},
archivePrefix = {arXiv},
       eprint = {1811.10549},
 primaryClass = {hep-ph},
       adsurl = {https://ui.adsabs.harvard.edu/abs/2019PhRvD..99d3541E}
}

@ARTICLE{2020PhLB..80335325B,
       author = {{Baum}, Sebastian and {Drukier}, Andrzej K. and {Freese}, Katherine and {G{\'o}rski}, Maciej and {Stengel}, Patrick},
        title = "{Searching for dark matter with paleo-detectors}",
      journal = {Physics Letters B},
         year = 2020,
        month = apr,
       volume = {803},
          eid = {135325},
        pages = {135325},
          doi = {10.1016/j.physletb.2020.135325},
       adsurl = {https://ui.adsabs.harvard.edu/abs/2020PhLB..80335325B}
}

@ARTICLE{2021PhRvD.104l3015B,
       author = {{Baum}, Sebastian and {DeRocco}, William and {Edwards}, Thomas D.~P. and {Kalia}, Saarik},
        title = "{Galactic geology: Probing time-varying dark matter signals with paleodetectors}",
      journal = {\prd},
         year = 2021,
        month = dec,
       volume = {104},
       number = {12},
          eid = {123015},
        pages = {123015},
          doi = {10.1103/PhysRevD.104.123015},
archivePrefix = {arXiv},
       eprint = {2107.02812},
 primaryClass = {astro-ph.GA},
       adsurl = {https://ui.adsabs.harvard.edu/abs/2021PhRvD.104l3015B}
}

@ARTICLE{2021PASJ...73..639N,
       author = {{Nakazato}, Ken'ichiro and {Sumiyoshi}, Kohsuke and {Togashi}, Hajime},
        title = "{Numerical study of stellar core collapse and neutrino emission using the nuclear equation of state obtained by the variational method}",
      journal = {\pasj},
         year = 2021,
        month = jun,
       volume = {73},
       number = {3},
        pages = {639-651},
          doi = {10.1093/pasj/psab026},
archivePrefix = {arXiv},
       eprint = {2103.14386},
 primaryClass = {astro-ph.HE},
       adsurl = {https://ui.adsabs.harvard.edu/abs/2021PASJ...73..639N}
}

@ARTICLE{2022ApJ...925...98N,
       author = {{Nakazato}, Ken'ichiro and {Nakanishi}, Fumi and {Harada}, Masayuki and {Koshio}, Yusuke and {Suwa}, Yudai and {Sumiyoshi}, Kohsuke and {Harada}, Akira and {Mori}, Masamitsu and {Wendell}, Roger A.},
        title = "{Observing Supernova Neutrino Light Curves with Super-Kamiokande. II. Impact of the Nuclear Equation of State}",
      journal = {\apj},
         year = 2022,
        month = jan,
       volume = {925},
       number = {1},
          eid = {98},
        pages = {98},
          doi = {10.3847/1538-4357/ac3ae2},
archivePrefix = {arXiv},
       eprint = {2108.03009},
 primaryClass = {astro-ph.HE},
       adsurl = {https://ui.adsabs.harvard.edu/abs/2022ApJ...925...98N}
}

@ARTICLE{2009A&A...498...95V,
       author = {{Vanhollebeke}, E. and {Groenewegen}, M.~A.~T. and {Girardi}, L.},
        title = "{Stellar populations in the Galactic bulge. Modelling the Galactic bulge with TRILEGAL}",
      journal = {\aap},
         year = 2009,
        month = apr,
       volume = {498},
       number = {1},
        pages = {95-107},
          doi = {10.1051/0004-6361/20078472},
archivePrefix = {arXiv},
       eprint = {0903.0946},
 primaryClass = {astro-ph.GA},
       adsurl = {https://ui.adsabs.harvard.edu/abs/2009A&A...498...95V}
}

@ARTICLE{2009MNRAS.398..263M,
       author = {{Majaess}, D.~J. and {Turner}, D.~G. and {Lane}, D.~J.},
        title = "{Characteristics of the Galaxy according to Cepheids}",
      journal = {\mnras},
         year = 2009,
        month = sep,
       volume = {398},
       number = {1},
        pages = {263-270},
          doi = {10.1111/j.1365-2966.2009.15096.x},
archivePrefix = {arXiv},
       eprint = {0903.4206},
 primaryClass = {astro-ph.GA},
       adsurl = {https://ui.adsabs.harvard.edu/abs/2009MNRAS.398..263M}
}

@ARTICLE{2014MNRAS.441.1105F,
       author = {{Francis}, Charles and {Anderson}, Erik},
        title = "{Two estimates of the distance to the Galactic Centre}",
      journal = {\mnras},
         year = 2014,
        month = jun,
       volume = {441},
       number = {2},
        pages = {1105-1114},
          doi = {10.1093/mnras/stu631},
archivePrefix = {arXiv},
       eprint = {1309.2629},
 primaryClass = {astro-ph.GA},
       adsurl = {https://ui.adsabs.harvard.edu/abs/2014MNRAS.441.1105F}
}

@ARTICLE{2019A&A...625L..10G,
       author = {{GRAVITY Collaboration} and {Abuter}, R. and {Amorim}, A. and {Baub{\"o}ck}, M. and {Berger}, J.~P. and {Bonnet}, H. and {Brandner}, W. and {Cl{\'e}net}, Y. and {Coud{\'e} Du Foresto}, V. and {de Zeeuw}, P.~T. and {Dexter}, J. and {Duvert}, G. and {Eckart}, A. and {Eisenhauer}, F. and {F{\"o}rster Schreiber}, N.~M. and {Garcia}, P. and {Gao}, F. and {Gendron}, E. and {Genzel}, R. and {Gerhard}, O. and {Gillessen}, S. and {Habibi}, M. and {Haubois}, X. and {Henning}, T. and {Hippler}, S. and {Horrobin}, M. and {Jim{\'e}nez-Rosales}, A. and {Jocou}, L. and {Kervella}, P. and {Lacour}, S. and {Lapeyr{\`e}re}, V. and {Le Bouquin}, J.-B. and {L{\'e}na}, P. and {Ott}, T. and {Paumard}, T. and {Perraut}, K. and {Perrin}, G. and {Pfuhl}, O. and {Rabien}, S. and {Rodriguez Coira}, G. and {Rousset}, G. and {Scheithauer}, S. and {Sternberg}, A. and {Straub}, O. and {Straubmeier}, C. and {Sturm}, E. and {Tacconi}, L.~J. and {Vincent}, F. and {von Fellenberg}, S. and {Waisberg}, I. and {Widmann}, F. and {Wieprecht}, E. and {Wiezorrek}, E. and {Woillez}, J. and {Yazici}, S.},
        title = "{A geometric distance measurement to the Galactic center black hole with 0.3\% uncertainty}",
      journal = {\aap},
         year = 2019,
        month = may,
       volume = {625},
          eid = {L10},
        pages = {L10},
          doi = {10.1051/0004-6361/201935656},
archivePrefix = {arXiv},
       eprint = {1904.05721},
 primaryClass = {astro-ph.GA},
       adsurl = {https://ui.adsabs.harvard.edu/abs/2019A&A...625L..10G}
}

@ARTICLE{2020PASJ...72...50V,
       author = {{VERA Collaboration} and {Hirota}, Tomoya and {Nagayama}, Takumi and {Honma}, Mareki and {Adachi}, Yuuki and {Burns}, Ross A. and {Chibueze}, James O. and {Choi}, Yoon Kyung and {Hachisuka}, Kazuya and {Hada}, Kazuhiro and {Hagiwara}, Yoshiaki and {Hamada}, Shota and {Handa}, Toshihiro and {Hashimoto}, Mao and {Hirano}, Ken and {Hirata}, Yushi and {Ichikawa}, Takanori and {Imai}, Hiroshi and {Inenaga}, Daichi and {Ishikawa}, Toshio and {Jike}, Takaaki and {Kameya}, Osamu and {Kaseda}, Daichi and {Kim}, Jeong Sook and {Kim}, Jungha and {Kim}, Mi Kyoung and {Kobayashi}, Hideyuki and {Kono}, Yusuke and {Kurayama}, Tomoharu and {Matsuno}, Masako and {Morita}, Atsushi and {Motogi}, Kazuhito and {Murase}, Takeru and {Nakagawa}, Akiharu and {Nakanishi}, Hiroyuki and {Niinuma}, Kotaro and {Nishi}, Junya and {Oh}, Chung Sik and {Omodaka}, Toshihiro and {Oyadomari}, Miyako and {Oyama}, Tomoaki and {Sakai}, Daisuke and {Sakai}, Nobuyuki and {Sawada-Satoh}, Satoko and {Shibata}, Katsunori M. and {Shizugami}, Makoto and {Sudo}, Jumpei and {Sugiyama}, Koichiro and {Sunada}, Kazuyoshi and {Suzuki}, Syunsaku and {Takahashi}, Ken and {Tamura}, Yoshiaki and {Tazaki}, Fumie and {Ueno}, Yuji and {Uno}, Yuri and {Urago}, Riku and {Wada}, Koji and {Wu}, Yuan Wei and {Yamashita}, Kazuyoshi and {Yamashita}, Yuto and {Yamauchi}, Aya and {Yuda}, Akito},
        title = "{The First VERA Astrometry Catalog}",
      journal = {\pasj},
         year = 2020,
        month = aug,
       volume = {72},
       number = {4},
          eid = {50},
        pages = {50},
          doi = {10.1093/pasj/psaa018},
archivePrefix = {arXiv},
       eprint = {2002.03089},
 primaryClass = {astro-ph.GA},
       adsurl = {https://ui.adsabs.harvard.edu/abs/2020PASJ...72...50V}
}

@ARTICLE{2008ApJ...684.1336K,
       author = {{Kochanek}, Christopher S. and {Beacom}, John F. and {Kistler}, Matthew D. and {Prieto}, Jos{\'e} L. and {Stanek}, Krzysztof Z. and {Thompson}, Todd A. and {Y{\"u}ksel}, Hasan},
        title = "{A Survey About Nothing: Monitoring a Million Supergiants for Failed Supernovae}",
      journal = {\apj},
         year = 2008,
        month = sep,
       volume = {684},
       number = {2},
        pages = {1336-1342},
          doi = {10.1086/590053},
archivePrefix = {arXiv},
       eprint = {0802.0456},
 primaryClass = {astro-ph},
       adsurl = {https://ui.adsabs.harvard.edu/abs/2008ApJ...684.1336K}
}

@ARTICLE{2010NIMPB.268.1818Z,
       author = {{Ziegler}, James F. and {Ziegler}, M.~D. and {Biersack}, J.~P.},
        title = "{SRIM - The stopping and range of ions in matter (2010)}",
      journal = {Nuclear Instruments and Methods in Physics Research B},
         year = 2010,
        month = jun,
       volume = {268},
       number = {11-12},
        pages = {1818-1823},
          doi = {10.1016/j.nimb.2010.02.091},
       adsurl = {https://ui.adsabs.harvard.edu/abs/2010NIMPB.268.1818Z}
}

@ARTICLE{2025PhRvD.112d3040F,
       author = {{Fung}, Audrey and {Lucas}, Thalles and {Balogh}, Levente and {Leybourne}, Matthew and {Vincent}, Aaron C.},
        title = "{Refining the sensitivity of new physics searches with ancient minerals}",
      journal = {\prd},
         year = 2025,
        month = aug,
       volume = {112},
       number = {4},
          eid = {043040},
        pages = {043040},
          doi = {10.1103/clvr-mhx5},
archivePrefix = {arXiv},
       eprint = {2504.08885},
 primaryClass = {hep-ph},
       adsurl = {https://ui.adsabs.harvard.edu/abs/2025PhRvD.112d3040F}
}

@TECHREPORT{Madland1999-ee,
  title     = "{SOURCES} 4A: A code for calculating (alpha,n), spontaneous
               fission, and delayed neutron sources and spectra",
  author    = "Madland, D G and Arthur, E D and Estes, G P and Stewart, J E and
               Bozoian, M and Perry, R T and Parish, T A and Brown, T H and
               England, T R and Wilson, W B and Charlton, W S",
  publisher = "Office of Scientific and Technical Information (OSTI)",
  month     =  sep,
  year      =  1999,
  doi       =  "10.2172/15215"
}

@ARTICLE{2014NDS...120..294S,
       author = {{Soppera}, N. and {Bossant}, M. and {Dupont}, E.},
        title = "{JANIS 4: An Improved Version of the NEA Java-based Nuclear Data Information System}",
      journal = {Nuclear Data Sheets},
         year = 2014,
        month = jun,
       volume = {120},
        pages = {294-296},
          doi = {10.1016/j.nds.2014.07.071},
       adsurl = {https://ui.adsabs.harvard.edu/abs/2014NDS...120..294S}
}

@ARTICLE{2016PhRvD..94f3527O,
       author = {{O'Hare}, Ciaran A.~J.},
        title = "{Dark matter astrophysical uncertainties and the neutrino floor}",
      journal = {\prd},
         year = 2016,
        month = sep,
       volume = {94},
       number = {6},
          eid = {063527},
        pages = {063527},
          doi = {10.1103/PhysRevD.94.063527},
archivePrefix = {arXiv},
       eprint = {1604.03858},
 primaryClass = {astro-ph.CO},
       adsurl = {https://ui.adsabs.harvard.edu/abs/2016PhRvD..94f3527O}
}

@ARTICLE{2020PhRvL.125w1802J,
       author = {{Jordan}, Johnathon R. and {Baum}, Sebastian and {Stengel}, Patrick and {Ferrari}, Alfredo and {Morone}, Maria Cristina and {Sala}, Paola and {Spitz}, Joshua},
        title = "{Measuring Changes in the Atmospheric Neutrino Rate over Gigayear Timescales}",
      journal = {\prl},
         year = 2020,
        month = dec,
       volume = {125},
       number = {23},
          eid = {231802},
        pages = {231802},
          doi = {10.1103/PhysRevLett.125.231802},
archivePrefix = {arXiv},
       eprint = {2004.08394},
 primaryClass = {hep-ph},
       adsurl = {https://ui.adsabs.harvard.edu/abs/2020PhRvL.125w1802J}
}

@ARTICLE{2021PhRvD.103l3016T,
       author = {{Tapia-Arellano}, Natalia and {Horiuchi}, Shunsaku},
        title = "{Measuring solar neutrinos over gigayear timescales with paleo detectors}",
      journal = {\prd},
         year = 2021,
        month = jun,
       volume = {103},
       number = {12},
          eid = {123016},
        pages = {123016},
          doi = {10.1103/PhysRevD.103.123016},
archivePrefix = {arXiv},
       eprint = {2102.01755},
 primaryClass = {hep-ph},
       adsurl = {https://ui.adsabs.harvard.edu/abs/2021PhRvD.103l3016T}
}

@ARTICLE{2010ARNPS..60..439B,
       author = {{Beacom}, John F.},
        title = "{The Diffuse Supernova Neutrino Background}",
      journal = {Annual Review of Nuclear and Particle Science},
         year = 2010,
        month = nov,
       volume = {60},
        pages = {439-462},
          doi = {10.1146/annurev.nucl.010909.083331},
archivePrefix = {arXiv},
       eprint = {1004.3311},
 primaryClass = {astro-ph.HE},
       adsurl = {https://ui.adsabs.harvard.edu/abs/2010ARNPS..60..439B}
}

@ARTICLE{2014NatSR...4.3857H,
       author = {{Holler}, M. and {Diaz}, A. and {Guizar-Sicairos}, M. and {Karvinen}, P. and {F{\"a}rm}, Elina and {H{\"a}rk{\"o}nen}, Emma and {Ritala}, Mikko and {Menzel}, A. and {Raabe}, J. and {Bunk}, O.},
        title = "{X-ray ptychographic computed tomography at 16{\hsapce{0.25em}}nm isotropic 3D resolution}",
      journal = {Scientific Reports},
         year = 2014,
        month = jan,
       volume = {4},
          eid = {3857},
        pages = {3857},
          doi = {10.1038/srep03857},
       adsurl = {https://ui.adsabs.harvard.edu/abs/2014NatSR...4.3857H}
}

@ARTICLE{2026Sci...391..689D,
       author = {{De}, Kishalay and {MacLeod}, Morgan and {Jencson}, Jacob E. and {Lovegrove}, Elizabeth and {Antoni}, Andrea and {Kara}, Erin and {Kasliwal}, Mansi M. and {Lau}, Ryan M. and {Loeb}, Abraham and {Masterson}, Megan and {Meisner}, Aaron M. and {Panagiotou}, Christos and {Quataert}, Eliot and {Simcoe}, Robert},
        title = "{Disappearance of a massive star in the Andromeda Galaxy due to formation of a black hole}",
      journal = {Science},
         year = 2026,
        month = feb,
       volume = {391},
       number = {6786},
        pages = {689-693},
          doi = {10.1126/science.adt4853},
archivePrefix = {arXiv},
       eprint = {2410.14778},
 primaryClass = {astro-ph.HE},
       adsurl = {https://ui.adsabs.harvard.edu/abs/2026Sci...391..689D}
}

@ARTICLE{2022PhRvD.106l3008B,
       author = {{Baum}, Sebastian and {Capozzi}, Francesco and {Horiuchi}, Shunsaku},
        title = "{Rocks, water, and noble liquids: Unfolding the flavor contents of supernova neutrinos}",
      journal = {\prd},
         year = 2022,
        month = dec,
       volume = {106},
       number = {12},
          eid = {123008},
        pages = {123008},
          doi = {10.1103/PhysRevD.106.123008},
archivePrefix = {arXiv},
       eprint = {2203.12696},
 primaryClass = {hep-ph},
       adsurl = {https://ui.adsabs.harvard.edu/abs/2022PhRvD.106l3008B}
}

@misc{yamasaki_2026_21002077,
  author       = {Yamasaki, Mahiro and
                  Nakazato, Ken'ichiro},
  title        = {Paleodetectors for Neutrino Signals from Diverse
                   Stellar Collapses in the Galaxy
                  },
  month        = jul,
  year         = 2026,
  publisher    = {Zenodo},
  doi          = {10.5281/zenodo.21002077},
  url          = {https://doi.org/10.5281/zenodo.21002077},
}

\end{document}